\documentclass[twocolumn]{aastex7}
\usepackage{CJK}
\usepackage{enumitem}
\usepackage{amsmath}
\usepackage{siunitx}
\usepackage{soul}
\usepackage{textcomp}

\newcommand{\sqdeg}{\mathrm{deg}^2}
\newcommand{\pqso}{p_{\mathrm{QSO}}}
\shorttitle{QHSC: The HSC Quasar Candidate Catalog}
\shortauthors{Zhu et al.}
\graphicspath{{./}}

\begin{document}
\begin{CJK*}{UTF8}{gbsn}

\title{QHSC: The Quasar Candidate Catalog for the Hyper Suprime-Cam Subaru Strategic Program}

\correspondingauthor{Xue-Bing Wu}

\author[orcid=0000-0002-0792-2353, sname='Zhu', gname='Rui']{Rui Zhu (朱睿)}
\affiliation{Department of Astronomy, School of Physics, Peking University, Beijing 100871, People's Republic of China}
\affiliation{Kavli Institute for Astronomy and Astrophysics, Peking University, Beijing 100871, People's Republic of China}
\email[show]{rui@stu.pku.edu.cn}

\author[orcid=0000-0002-7350-6913, sname='Wu', gname='Xue-Bing']{Xue-Bing Wu (吴学兵)}
\affiliation{Department of Astronomy, School of Physics, Peking University, Beijing 100871, People's Republic of China}
\affiliation{Kavli Institute for Astronomy and Astrophysics, Peking University, Beijing 100871, People's Republic of China}
\email[show]{wuxb@pku.edu.cn}

\author[orcid=0009-0005-3823-9302, sname='Pang', gname='Yuxuan']{Yuxuan Pang (庞宇萱)}
\affiliation{Department of Astronomy, School of Physics, Peking University, Beijing 100871, People's Republic of China}
\affiliation{Kavli Institute for Astronomy and Astrophysics, Peking University, Beijing 100871, People's Republic of China}
\affiliation{School of Astronomy and Space Science, University of Chinese Academy of Sciences (UCAS), Beijing 100049, China}
\email{jackdaw.pyx@gmail.com}

\author[orcid=0000-0002-0759-0504, sname='Fu', gname='Yuming']{Yuming Fu (傅煜铭)}
\affiliation{Leiden Observatory, Leiden University, Einsteinweg 55, 2333 CC Leiden, The Netherlands}
\affiliation{Kapteyn Astronomical Institute, University of Groningen, P.O. Box 800, 9700 AV Groningen, The Netherlands}
\email{yfu@strw.leidenuniv.nl}

\begin{abstract}
The Hyper Suprime-Cam Subaru Strategic Program (HSC-SSP) is a deep wide-field multi-band imaging survey consisting of three layers (Wide, Deep, and UltraDeep), with the Wide layer covering $\sim 1470$ deg$^2$ to a depth of $i \sim 26$ mag. We present the QHSC catalog, a machine-learning selected sample of quasar candidates with photometric redshifts in the Wide layer of the HSC-SSP survey (Public Data Release 3). The full QHSC catalog contains four distinct samples: a master sample with HSC-only photometry, an HSC+WISE sample, and two samples including near-infrared data from UKIDSS and VISTA, denoted as GoldenU and GoldenV. For each sample, an XGBoost classifier is trained and evaluated using independent spectroscopic test sets from HETDEX, VVDS, and zCOSMOS-bright. The numbers of quasar candidates in the QHSC catalog are 1,184,574 (master), 371,777 (HSC+WISE), 87,460 (GoldenU), and 120,572 (GoldenV), with respective completeness values of 85.3\%, 92.7\%, 89.8\%, and 91.3\%. We develop ensemble photometric redshift estimators based on bootstrap aggregating (bagging) of multiple XGBoost regressors, achieving outlier fractions of 21.7\%, 13.1\%, 9.5\%, and 10.7\% for these samples. The catalog provides quasar classification probabilities ($\pqso$), enabling construction of purer subsamples via thresholding. This work offers a valuable resource for studies of quasars and cosmology, and highlights the effectiveness of machine learning for quasar selection in future wide and deep imaging surveys. The catalog is publicly available at \url{https://doi.org/10.5281/zenodo.17515028}.
\end{abstract}



\section{Introduction} \label{sec:intro}
Quasars (QSOs) are a type of luminous active galactic nuclei (AGNs) powered by supermassive black holes (SMBHs) at their centers. They emit enormous amounts of energy through intense accretion processes, shining across vast cosmic distances and making them ideal probes for various astrophysical and cosmological studies \citep{1984ARA&A..22..471R}. For instance, the discovery of quasars with especially massive black holes in the early universe challenges our understanding of SMBH growth \citep{2015Natur.518..512W,2023ARA&A..61..373F,2020ARA&A..58...27I}. Furthermore, the statistics of a quasar sample are used to investigate the co-evolution of SMBHs and host galaxies \citep{2013ARA&A..51..511K,2005Natur.433..604D}.

The Million Quasars catalogue \citep[Milliquas v8; ][]{2023OJAp....6E..49F} has been the largest compilation of quasars, incorporating discoveries from various published works. The final release includes all quasars published up to 30 June 2023, with nearly one million sources identified, the majority of which have been spectroscopically confirmed by the Sloan Digital Sky Survey \citep[SDSS;][]{2020ApJS..250....8L}. Meanwhile, ongoing spectroscopic programs, such as the Dark Energy Spectroscopic Instrument \citep[DESI; ][]{2024AJ....168...58D} and the LAMOST quasar survey \citep{2023ApJS..265...25J}, have contributed significantly to the growing number of spectroscopically confirmed quasars. However, due to the limited depth and observational efficiency of spectroscopic surveys, the confirmed quasar sample remains incomplete.

Photometric surveys, which cover vastly more sources than spectroscopic surveys, have greatly expanded the pool of quasar candidates, motivating the development of efficient selection methods. These methods primarily exploit the intrinsic differences in the spectral energy distributions (SEDs) of quasars and other celestial objects, which are reflected in the color space constructed from multi-band photometric data. Traditional techniques often employ simple color cuts in two or more bands to separate quasars from non-quasar sources \citep[e.g.][]{1999AJ....117.2528F,2004ApJS..155..257R,2012AJ....144...49W,2016ApJ...833..222J,2016ApJ...819...24W}. More recently, machine learning (ML) approaches have been increasingly applied to efficiently identify quasar candidates in high-dimensional parameter spaces using wide-area photometric and astrometric data \citep[e.g.][]{2004A&A...422.1113Z,2009RAA.....9..220G,2012MNRAS.425.2599P,2018A&A...619A..14F,2021MNRAS.501.3951C,2024MNRAS.527.4677Z,2024ApJS..271...54F,2025ApJS..279...54F}.

The Hyper Suprime-Cam Subaru Strategic Program \citep[HSC-SSP;][]{2018PASJ...70S...4A,2022PASJ...74..247A}, conducted with the Subaru 8.2\,m telescope, represents the state-of-the-art deep and wide-area optical to near-infrared imaging survey. Several efforts have been made to identify quasars within the HSC-SSP fields, particularly focusing on faint, high-redshift populations. Notably, the Subaru High-$z$ Exploration of Low-Luminosity Quasars (SHELLQs) project, has spectroscopically confirmed 162 low-luminosity quasars at $z > 5.6$ \citep{2016ApJ...828...26M,2022ApJS..259...18M}, enabling robust constraints on the faint-end quasar luminosity function (QLF) and revealing a significant flattening at $z \sim 6$ toward lower luminosities. 

Recently, the superior angular resolution, high-quality point spread function (PSF), and unprecedented infrared sensitivity of the James Webb Space Telescope \citep[JWST;][]{2023PASP..135d8001R} have enabled detailed studies of SHELLQs quasar host galaxies during the epoch of reionization ($z > 6$). These observations provide critical insights into the early growth of SMBHs and the co-evolution of galaxies and their central SMBHs \citep{2023Natur.621...51D, 2025arXiv250503876D,2025ApJ...988...57M}. Additionally, \citet{2021MNRAS.503.4136G} demonstrated the effectiveness of the gradient-boosted decision trees (GBDT) algorithm in classifying galaxies, stars, and AGNs in deep imaging surveys, using HSC-SSP data supplemented with $U$-band photometry from the CLAUDS survey \citep{2019MNRAS.489.5202S}. Although the combined dataset is exceptionally deep, it covers a relatively small area of approximately $18.6~\sqdeg$ within the HSC-SSP Deep and UltraDeep fields.

In this paper, we perform quasar candidate selection and photometric redshift (photo-$z$) estimation in the HSC-SSP Wide fields using machine learning methods. In particular, we evaluate the classification performance using independent test sets from three deep spectroscopic surveys: the VIMOS VLT Deep Survey \citep[VVDS;][]{2013A&A...559A..14L}, zCOSMOS-bright survey \citep{2012ApJ...753..121K}, and the Hobby-Eberly Telescope Dark Energy Experiment survey \citep[HETDEX;][]{2023ApJ...943..177M}. The XGBoost algorithm \citep{2016arXiv160302754C}, a state-of-the-art implementation of the GBDT algorithm, is employed in this work. Previous studies have demonstrated that XGBoost achieves high purity and completeness for quasar selection, outperforming other ML algorithms such as random forests and neural networks \citep{2021A&A...649A..81N,2024ApJS..271...54F}. To further improve the classification and photo-$z$ accuracy, we construct four parent samples with varying depths and wavelength coverage of photometric bands by incorporating a combination of near-infrared (near-IR) broad-band photometry from the UKIRT Infrared Deep Sky Survey \citep[UKIDSS;][]{2007MNRAS.379.1599L}, the VISTA Kilo-degree Infrared Galaxy Survey \citep[VIKING;][]{2013Msngr.154...32E}, the VISTA Hemisphere Survey \citep[VHS;][]{2013Msngr.154...35M}, and mid-infrared (mid-IR) broad-band photometry from the CatWISE2020 catalog \citep{2021ApJS..253....8M}. We combine the results for all parent samples as the QHSC quasar candidate catalog.

The paper is organized as follows. Section~\ref{sec:data} describes the photometric data used to construct the parent samples, as well as the spectroscopic data employed for training and testing. In Section~\ref{sec:ml_selection}, we detail the machine learning classification pipeline, which includes feature selection, model optimization, and evaluation of selection efficiency. Section~\ref{sec:photoz} presents our photo-$z$ estimation method based on ensemble learning for each quasar candidate sample. In Section~\ref{sec:uband}, we evaluate the impact of including external $u$-band photometry on both classification and photo-$z$ estimation. Finally, Section~\ref{sec:summary} summarizes the main results and conclusions. Throughout this paper, we adopt a flat $\Lambda$CDM cosmology with parameters $\Omega_{\Lambda} = 0.7$, $\Omega_{\rm M} = 0.3$, and $H_0 = 70\;\mathrm{km\;s^{-1}\;Mpc^{-1}}$.

\section{Data} \label{sec:data}

\begin{figure*}[ht]
    \centering
    \includegraphics[width=1\textwidth]{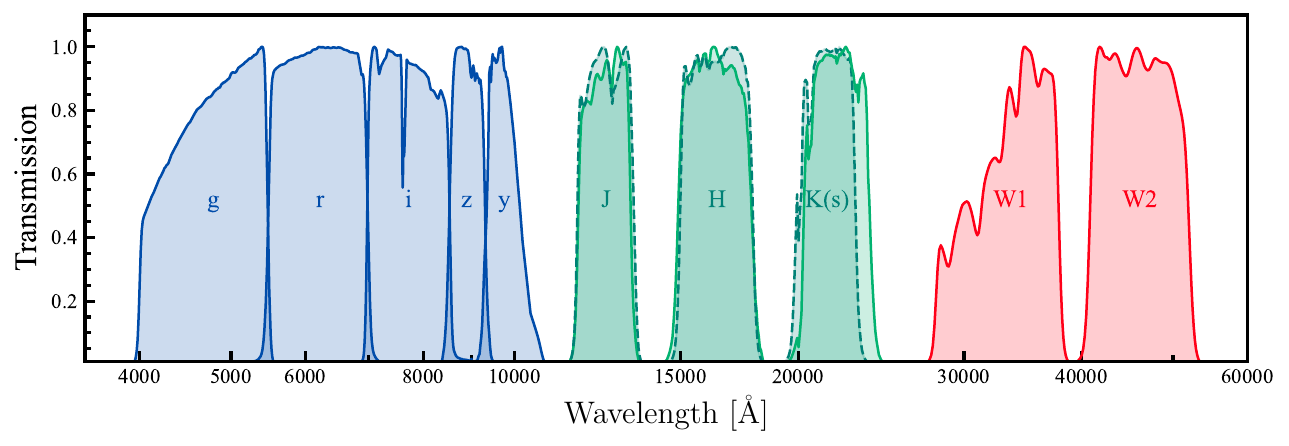}
    \caption{
    The full set of filter transmission curves used in this work, normalized to a maximum transmission of one. Blue and red lines represent the HSC and WISE filters, respectively. Cyan lines represent near-infrared filters from UKIDSS (solid) and VISTA (dashed).
    }
    \label{fig:filter_trans}
\end{figure*}

Our analysis is based on the forced photometric catalog (\texttt{pdr3\_wide.forced} table) from the Wide layer of the HSC-SSP Public Data Release 3 \citep[][]{2022PASJ...74..247A}. After applying a series of quality selection criteria, we cross-match the HSC sources with near-IR data from the UKIDSS survey and four VISTA survey programs (VIKING, VHS, UltraVISTA, and VIDEO), as well as mid-IR data from the WISE survey, using a matching radius of $1''$ adopted throughout this work. Based on these photometric datasets, we construct four parent samples: the master sample, the HSC+WISE sample, the GoldenU sample, and the GoldenV sample. Figure~\ref{fig:filter_trans} shows the optical to mid-IR filter sets used for quasar selection and photo-$z$ estimation in each parent sample. The master sample includes HSC Wide-layer sources with reliable HSC photometry, covering wavelengths from 400 to 1070\,nm, selected using conservative quality cuts (see Section~\ref{subsubsec:HSC_data} for details on photometric data cleaning and Section~\ref{subsec:parent_samples} for the construction of the parent samples). Sources within the master sample that have reliable WISE W1 and W2 counterparts are extracted to form the HSC+WISE sample. The golden samples further incorporate near-IR photometry on top of the HSC+WISE sample, achieving a more complete wavelength coverage from 400\,nm to approximately 5\,$\mu$m. Nevertheless, the golden samples are shallower and cover a smaller area than the master and HSC+WISE samples. Specifically, we use UKIDSS $JHK$ data to construct the GoldenU sample and VISTA $JHK_\mathrm{s}$ data to construct the GoldenV sample, depending on survey availability in different HSC fields. Spectroscopic classifications from SDSS and DESI DR1 are used to build the training sets, while additional spectra from VVDS, zCOSMOS-bright, and HETDEX are incorporated to construct well-defined test sets.

\subsection{Photometric Data}
\subsubsection{The HSC Data} \label{subsubsec:HSC_data}

The HSC-SSP survey consists of three layers: Wide, Deep, and UltraDeep, each with a distinct survey area and depth. The Wide layer covers an area of approximately $1470~\sqdeg$, primarily along the celestial equator, using five broadband filters ($g$, $r$, $i$, $z$, and $y$). The targeted $5\sigma$ limiting magnitudes in $2''$ diameter apertures are ($g$, $r$, $i$, $z$, $y$) = (26.5, 26.1, 25.9, 25.1, 24.4) AB mag. The Deep and UltraDeep layers achieve greater depths of approximately 26.9 mag and 27.7 mag, respectively, while covering smaller sky areas of $26~\sqdeg$ and $3.5~\sqdeg$. This unprecedented combination of depth and area enables detailed investigations of quasar properties across a wide range of brightness and redshift.

The HSC-SSP PDR3 data were processed using \texttt{hscPipe} v8 \citep{2018PASJ...70S...5B}, a customized version of the science pipelines developed for the Rubin Observatory Legacy Survey of Space and Time \citep[LSST;][]{2019ApJ...873..111I}. The forced photometry catalog provides celestial coordinates, broadband photometry, and data quality flags for 507,215,729 unique sources (\texttt{isprimary = True}). We select three types of photometry from the PDR3 catalog: point-spread function fitting (PSF) photometry, CModel photometry (a detailed description of the HSC CModel algorithm is provided in Appendix~2 of \citealt{2018PASJ...70S...5B}), and $2''$ aperture photometry. Photometric measurements for all unique sources are retrieved from the catalog. We apply two magnitude corrections based on the calibration tables \texttt{pdr3\_wide.magcorr} and \texttt{pdr3\_wide.stellar\_sequence\_offsets}. The former accounts for differences between the original HSC $r/i$ bands and the updated $r2/i2$ bands, while the latter offset is used to correct the photometric zero-point for each patch\footnote{A subregion in HSC images introduced to parallelize the processing, serving as the most fundamental areal unit.}, as estimated from the location of the stellar sequence \citep[see ][]{2022PASJ...74..247A}.

\begin{deluxetable}{lc}
\tablecaption{\centering Flag values and their meanings for image and photometry. \label{tab:new_flags}}
\tabletypesize{\footnotesize}
\tablehead{
\colhead{Flag Value} & \colhead{Meaning}
}
\startdata
\multicolumn{2}{c}{\textbf{Image Flag}} \\
\hline
\texttt{image\_flag = 0} & HSC image is ideal \\
\texttt{image\_flag = 1} & HSC image is absent \\
\texttt{image\_flag = 2} & HSC image has an error \\
\hline
\multicolumn{2}{c}{\textbf{Photometry Flag}} \\
\hline
\texttt{phot\_flag = 0} & Photometry SNR $\geq$ 3 and unsaturated \\
\texttt{phot\_flag = 1} & Photometry is missing due to image problem \\
\texttt{phot\_flag = 2} & Photometric SNR $<$ 3 \\
\texttt{phot\_flag = 3} & Photometry is potentially saturated \\
\enddata
\end{deluxetable}

In the data cleaning process, we utilize the data quality flags from the PDR3 catalog to define two general-purpose flags: an image flag and a photometry flag, which indicate the data quality of imaging and photometry in each HSC band, respectively. Table~\ref{tab:new_flags} summarizes the definitions of these newly introduced flags. A source is assigned \texttt{image\_flag = 1} in a given band if no HSC image is available. To consider an image as valid (\texttt{image\_flag = 0}), all of the following conditions must be simultaneously satisfied: 

\begin{itemize}
    \setlength\itemindent{0pt}
    \setlength\leftskip{0pt}
    \renewcommand\labelitemi{}
    \item \texttt{[grizy]\_pixelflags\_edge = False}
    \item \texttt{[grizy]\_pixelflags\_crcenter = False}
    \item \texttt{[grizy]\_pixelflags\_saturatedcenter = False}
    \item \texttt{[grizy]\_pixelflags\_interpolatedcenter = False}
\end{itemize}

\noindent Corresponding to these criteria, the source must be located within a usable exposure region, its center must be unaffected by cosmic rays, and the central pixel must be neither saturated nor interpolated. Otherwise, the image is flagged as anomalous (\texttt{image\_flag = 2}).

If the image in a given HSC band is considered valid (i.e., \texttt{image\_flag = 0}), we further evaluate the quality of the photometric measurements. Photometry in each band is flagged as (i) SNR $>$ 3 and unsaturated; (ii) low signal-to-noise ratio (SNR $<$ 3); or (iii) potentially saturated, based on the reference saturation magnitudes of $(g, r, i, z, y) = (17.4, 18.1, 18.3, 17.5, 17.0)$ reported by \citet{2022PASJ...74..247A}. Photometric values flagged by \texttt{phot\_flag} $\neq 0$ are treated as missing in subsequent analysis.

\subsubsection{The WISE data}
The Wide-field Infrared Survey Explorer \citep[WISE;][]{2010AJ....140.1868W} is a NASA Medium-Class Explorer mission that conducted a mid-infrared imaging survey of the entire sky in four bands centered at 3.4, 4.6, 12, and 22\,$\mu$m (denoted as W1, W2, W3, and W4, respectively). Building on data from the WISE and NEOWISE \citep{2011ApJ...731...53M} missions, the CatWISE2020 catalog \citep{2021ApJS..253....8M} provides nearly 1.9 billion sources across the entire sky in the W1 and W2 bands. The $5\sigma$ limiting magnitudes in the Vega system are W1 = 17.43 mag and W2 = 16.47 mag \citep{2021ApJS..253....8M}. Since the W1 and W2 magnitudes in the CatWISE2020 catalog are reported in the Vega system, we convert them to the AB system by adding an offset of 2.699 in W1, and 3.339 in W2 \citep{2010AJ....140.1868W}.

We cross-match the HSC coordinates with the CatWISE2020 catalog using a matching radius of $1''$ with the \texttt{STILTS} software \citep{2005ASPC..347...29T}. Due to the relatively low angular resolution of $\sim6''$ in the WISE W1 and W2 bands, cases where multiple HSC sources are matched to a single WISE source within the matching radius are excluded from the parent sample to avoid ambiguous associations ($\sim7.4\%$ of HSC–WISE pairs with separations $<1''$). Additional selection criteria are applied: (i) avoiding possible saturation by requiring \texttt{W1mproPM > 7} and \texttt{W2mproPM > 7}; and (ii) selecting sources significantly detected in both W1 and W2 bands with \texttt{snrW1pm > 5} and \texttt{snrW2pm > 5}.

\subsubsection{The UKIDSS data}
The UKIRT Infrared Deep Sky Survey \citep[UKIDSS;][]{2007MNRAS.379.1599L} was conducted using the Wide Field Camera (WFCAM) mounted on the 3.8-m United Kingdom Infrared Telescope (UKIRT), employing up to five filters ($Z$, $Y$, $J$, $H$, and $K$) spanning wavelengths from 0.83 to 2.37\,$\mu$m. The UKIDSS consists of five complementary surveys designed to probe different combinations of area, depth, and Galactic latitude. The Large Area Survey (LAS), the Galactic Cluster Survey (GCS), and the Galactic Plane Survey (GPS) collectively cover approximately $7000~\sqdeg$ to a depth of $K \sim 19.9$ mag (AB system). The Deep Extragalactic Survey (DXS) and the Ultra Deep Survey (UDS) reach depths of $K \sim 22.9$ mag and $K \sim 24.9$ mag, covering areas of $35~\sqdeg$ and $0.77~\sqdeg$, respectively.

The LAS, DXS, and UDS components of UKIDSS together cover a substantial portion of the northern part of the HSC fields. We select a sample of UKIDSS sources from these surveys that have detections in both the $J$ and $K$ bands with SNR $\geq 5$. To ensure unique and reliable detections, we apply the predicate (\texttt{priorsec = 0 OR priorsec = frameSetID}) to select sources that are either non-duplicates or the primary entries among duplicated sources \citep{2008MNRAS.384..637H}. We also apply the predicate (\texttt{mergedClass} $\neq$ 0 OR $-9$) to exclude noise and saturated detections. For the selected sources, we retrieve the $J$, $H$ (when available), and $K$ band $2''$ aperture magnitudes from the UKIDSS DR11PLUS\footnote{\url{http://wsa.roe.ac.uk/dr11plus_release.html}} catalog via the NOIRLab Astro Data Lab TAP service\footnote{\url{https://datalab.noirlab.edu/tap}}. All magnitudes are converted to the AB system using offsets of 0.938, 1.379, and 1.900 for the $J$, $H$, and $K$ bands, respectively, as given by \citet{2006MNRAS.367..454H}.

\subsubsection{The VISTA data}
The Visible and Infrared Survey Telescope for Astronomy \citep[VISTA;][]{2006Msngr.126...41E} is a 4-m class wide-field survey telescope equipped with a near-infrared camera ($\rm 0.85 \sim 2.3\,\mu m$), designed for extensive surveys of the Southern Hemisphere. Six major public surveys have been conducted with VISTA, each targeting different areas and depths, utilizing up to five filters ($Z$, $Y$, $J$, $H$, and $K_s$). These surveys include UltraVISTA \citep{2012A&A...544A.156M}, the VISTA Hemisphere Survey \citep[VHS;][]{2013Msngr.154...35M}, the VISTA Kilo-degree Infrared Galaxy Survey \citep[VIKING;][]{2013Msngr.154...32E}, the VISTA Variables in the Via Lactea Survey \citep[VVV;][]{2010NewA...15..433M}, the VISTA Deep Extragalactic Observations Survey \citep[VIDEO;][]{2013MNRAS.428.1281J}, and the VISTA Magellanic Survey \citep[VMC;][]{2011A&A...527A.116C}.

In this work, we utilize the VHS, VIKING, VIDEO, and UltraVISTA surveys as near-infrared data sets for the southern HSC fields. These surveys provide $5\sigma$ depths in the $K_s$ band (AB system) of approximately 20.0 mag (VHS), 21.2 mag (VIKING), 23.7 mag (VIDEO), and 24.0 mag (UltraVISTA). From these catalogs, we select sources detected in both the $J$ and $K_s$ bands with SNR $\geq 5$, and retrieve $J$, $H$ (when available), and $K_s$ band $2''$ aperture magnitudes. The VIKING DR4, VIDEO DR5, and UltraVISTA DR6 data are obtained via the ESO TAP service\footnote{\url{http://archive.eso.org/tap_cat}}, and the VHS DR5 data via the NOIRLab Astro Data Lab TAP. All magnitudes are converted from the Vega to the AB system using offsets of 0.916, 1.366, and 1.827 for the $J$, $H$, and $K_s$ bands, respectively, following \citet{2018MNRAS.474.5459G}.

\subsubsection{The SCUSS Data} \label{subsubsec:SCUSS}
The South Galactic Cap $u$-band Sky Survey \citep[SCUSS;][]{2016RAA....16...69Z, 2016AJ....151...37Z} is a deep $u$-band imaging survey covering approximately $5000~\sqdeg$ of the southern Galactic cap, conducted with the 2.3-m Bok Telescope operated by Steward Observatory. It substantially overlaps with the HSC Wide layer. The SCUSS $u_{\mathrm{AB}}$ filter is slightly bluer and narrower than the SDSS $u$-band, with a central wavelength of 3538\,\AA\ and a FWHM of 520\,\AA\ \citep{2015AJ....150..104Z}. The $5\sigma$ limiting magnitude of SCUSS reaches 23.2 mag in the $u_{\mathrm{AB}}$ band. We cross-match SCUSS sources to HSC with a radius of $1''$, and retrieve their PSF, CModel, and $2\farcs27$ aperture magnitudes measured from stacked images. To ensure reliable photometry, we apply the criteria SNR $\geq 3$ and \texttt{PSFFLAG} = 0 for each measurement. We use SCUSS $u$-band data to evaluate the performance of quasar selection and photo-$z$ estimation, as discussed in Section~\ref{sec:uband}.

\subsubsection{Galactic Extinction Correction} \label{subsubsec:extinction}
Extinction values from the optical to mid-infrared bands are computed using the SFD dust map \citep{1998ApJ...500..525S} and the extinction law of \citet{2007ApJ...663..320F}, adopting a fixed total-to-selective extinction ratio of $R_V = 3.1$. To account for the $\sim$14\% overestimation reported by \citet{2010ApJ...725.1175S}, we multiply a correction factor of 0.86 to the original SFD values. These extinction corrections are used in photometric redshift estimation and are not applied during classification.

\subsection{Spectroscopic Samples} \label{subsec:spec_samples}
Spectroscopic data from various surveys provide reliable classifications and accurate redshifts. We use spectroscopic samples from SDSS and DESI DR1 as the primary datasets for constructing the training set, which is used for both quasar selection and photometric redshift estimation. Due to its insufficient quality control, the Milliquas catalog is not used in building the quasar training set. In addition, we include a sample of very low-mass stars (VLMS), consisting of M, L, and T dwarfs and subdwarfs, in the training set because these objects can be easily confused with high-redshift or intrinsically red quasars \citep{2024ApJS..271...54F}. To evaluate the performance of our classification, we employ spectroscopic samples from three deep surveys: VVDS, zCOSMOS-bright, and HETDEX.

\subsubsection{The SDSS Survey}
The Sloan Digital Sky Survey \citep[SDSS;][]{2000AJ....120.1579Y} has mapped the high Galactic latitude northern sky through five-band photometry and spectroscopy of millions of objects, including galaxies, stars, and quasars. We utilize spectroscopic redshifts and classifications from two SDSS catalogs \citep{2022ApJS..259...35A,2023ApJS..267...44A}: the SDSS~I--IV SpecObj table\footnote{\url{https://dr17.sdss.org/sas/dr17/sdss/spectro/redux/specObj-dr17.fits}} and the SDSS-V SpAll (Run2d: v6\_0\_4) table\footnote{\url{https://dr18.sdss.org/sas/dr18/spectro/boss/redux/v6_0_4/spAll-v6_0_4.fits}}. These datasets are collectively referred to as the SpecObj sample. To ensure high-quality spectroscopic data, we apply the following selection criteria: \texttt{PLATEQUALITY = 'good' AND ZWARNING = 0 AND SPECPRIMARY = 1 AND SN\_MEDIAN\_ALL > 3}. To label the HSC sources as quasars, galaxies, and stars, we first select objects based on the \texttt{CLASS} parameter and then refine the classification using the \texttt{SUBCLASS} information. Quasars are defined as sources with \texttt{CLASS = 'QSO'} and \texttt{SUBCLASS} containing \texttt{BROADLINE}, while galaxies are defined as sources with \texttt{CLASS = 'GALAXY'} and \texttt{SUBCLASS} not containing \texttt{BROADLINE}.

In addition to the SpecObj sample, we incorporate quasars from the SDSS Quasar Catalog DR16Q \citep{2020ApJS..250....8L}, which contains 750{,}414 spectroscopically confirmed quasars, including 225{,}082 newly identified quasars in SDSS data releases. For these quasars, we adopt the systemic redshifts ($z_{\mathrm{sys}}$) reported by \citet{2022ApJS..263...42W}, which are derived from multiple emission lines and are more reliable than the original DR16Q redshifts ($z_{\mathrm{DR16Q}}$), especially at high redshift. We apply the following quality cuts to select robust DR16Q quasars: \texttt{Z\_SYS > 0 AND Z\_SYS\_ERR $\neq$ -1 AND Z\_SYS\_ERR $\neq$ -2}. For quasars appearing in both the SDSS SpecObj sample and the DR16Q catalog, we adopt the DR16Q entry. For quasars in the SpecObj sample that are not included in the DR16Q catalog, we compared their redshift estimates with those from the DESI DR1 catalog. The fraction of potentially problematic cases is small, with only $1.9\%$ showing $\Delta z / (1+z) > 0.15$.

\subsubsection{Data Release 1 of the DESI Survey}
The Dark Energy Spectroscopic Instrument Data Release 1 \citep[DESI DR1;][]{2025arXiv250314745D} comprises all data collected during the first 13 months of the DESI main survey, along with a uniform reprocessing of the DESI Survey Validation data. It represents the largest compilation of extragalactic redshifts to date. We select sources with high-confidence redshifts by applying the criteria \texttt{OBJTYPE = 'TGT' AND ZWARN = 0 AND ZCAT\_PRIMARY = True}, resulting in a sample of over 20 million objects. Based on spectroscopic classifications (Redrock spectral type), this sample includes 14,179,871 galaxies, 1,645,842 quasars, and 4,589,410 stars. To obtain more precise redshifts for DESI quasars, we adopt redshifts from FastSpecFit \citep{2023ascl.soft08005M}, as provided by the AGN/QSO Value-Added Catalog\footnote{\url{https://data.desi.lbl.gov/public/dr1/vac/dr1/agnqso}} for DESI DR1.

\subsubsection{The VVDS Survey}
The VIMOS VLT Deep Survey \citep[VVDS;][]{2004A&A...428.1043L, 2005A&A...439..845L, 2013A&A...559A..14L} is a magnitude-limited spectroscopic redshift survey conducted with the Visible Multi-Object Spectrograph (VIMOS) on the 8.2 m Melipal telescope of the European Southern Observatory’s Very Large Telescope (ESO VLT). VVDS targets are selected solely based on their $i$-band magnitude, reaching a depth of $i_{\mathrm{AB}} = 24.75$, without any color or morphological pre-selection. The survey consists of three complementary components: VVDS-Wide, VVDS-Deep, and VVDS-Ultra-Deep, covering sky areas of $8.7~\mathrm{deg}^2$, $0.74~\mathrm{deg}^2$, and $512~\mathrm{arcmin}^2$, respectively.

We combine data from all three VVDS components, yielding a total of 40,940 unique sources. These sources are used as a blind test sample to evaluate classification performance. Spectroscopic classifications and redshift quality are encoded in the \texttt{ZFLAGS} system, which is determined through visual inspection, where each spectrum is independently assessed by two VVDS team members \citep[see][]{2004A&A...428.1043L}. To construct a high-reliability ($\sim90\%$) test sample, we select sources with \texttt{ZFLAGS} values of 14, 214, 13, or 213 for the quasar sample, and 4, 24, 3, or 23 for the galaxy sample. The unit digit of the flag denotes the reliability, with a combination of 3 and 4 indicating a probability $>$95\% of being correct. A preceding digit further specifies additional information: "1" indicates that at least one emission line is broad, while "2" denotes that the object is not the primary target in the slit but falls within it by chance projection, thereby yielding a spectrum. Further details of the VVDS flag system are provided in Sec.~3.4 of \citet{2013A&A...559A..14L}.

\subsubsection{The zCOSMOS Survey}
The zCOSMOS survey \citep{2007ApJS..172...70L} is a large spectroscopic redshift survey in the COSMOS field, conducted with the VIMOS spectrograph on the 8.2 m ESO-VLT (Melipal telescope). It consists of two components: zCOSMOS-bright and zCOSMOS-deep. Similar to the VVDS survey, zCOSMOS-bright is a purely magnitude-limited survey targeting sources with $I_{\mathrm{AB}} < 22.5$, covering the entire $1.7~\mathrm{deg}^2$ COSMOS field. The zCOSMOS-deep survey focuses on approximately 10,000 galaxies selected by well-defined color criteria with $B_{\mathrm{AB}} < 25.25$, within the central $1~\mathrm{deg}^2$ region of COSMOS.

We adopt the zCOSMOS-bright sample as an additional test set. Similar to the \texttt{ZFLAGS} system used in VVDS, zCOSMOS-bright defines a ``Confidence Class (CC)'' flag system to quantify the reliability of individual redshifts and spectroscopic classifications \citep[see][]{2009ApJS..184..218L}. We construct the quasar test set using sources with the CC values 14.x, 13.x, 214.x, 213.x, 12.5, 212.5, and 18.5. The galaxy test set includes sources with CC values 4.x, 24.x, 3.x, 23.x, 2.5, 22.5, and 9.5. Here, ``x'' denotes any integer from 0 to 9. It is worth noting that zCOSMOS-bright excluded objects suspected to be Galactic stars (approximately 19\% of the photometric sample) from spectroscopic targeting, based on a combination of morphological and SED criteria. Consequently, the zCOSMOS-bright test sets are only applicable for validating quasar and galaxy classifications.

\subsubsection{The HETDEX Survey} 
The Hobby-Eberly Telescope Dark Energy Experiment \citep[HETDEX;][]{2021ApJ...923..217G} is a wide-area, untargeted spectroscopic survey conducted using the Visible Integral-field Replicable Unit Spectrograph \citep[VIRUS;][]{2010SPIE.7735E..3XL,2021AJ....162..298H} instrument on the 10~m Hobby-Eberly Telescope (HET), covering the wavelength range from 3500 to 5500~\AA. The survey spans a total area of $540~\sqdeg$. The first public data release, the HETDEX Public Source Catalog 1 \citep[hereafter SC1;][]{2023ApJ...943..177M}, includes 51,863 Ly$\alpha$-emitting galaxies (LAEs), 123,891 [O\,\textsc{ii}]-emitting galaxies, 37,916 stars, 5274 low-redshift galaxies without emission lines, and 4976 AGNs. To ensure high-confidence measurements, we select sources with \verb|z_hetdex_conf > 0.7|. The galaxy sample is restricted to objects with \texttt{source\_type} values of either \texttt{lzg} (low-redshift galaxies) or \texttt{oii} ([O\,\textsc{ii}]-emitting galaxies).

The HETDEX SC1 catalog is derived from the internal HETDEX Data Release (HDR2), which covers observations made between January 2017 and June 2020. In addition, a more recent AGN catalog \citep[e.g., the HETDEX HDR4 AGN catalog;][]{2025ApJS..276...72L} includes 15,940 AGNs across redshifts from $z=0.1$ to $z=4.6$ based on HETDEX HDR4, observed between January 2017 and August 2023. For consistency, we use the sources from HETDEX SC1 as the primary test set, while the new AGNs found in HETDEX HDR4 are exclusively used for photo-$z$ estimation.

\subsubsection{The VLMS sample}
The very low-mass stars (VLMS), including M, L, and T dwarfs as well as subdwarfs, are often misclassified as high-redshift or intrinsically red quasars due to their strong infrared emission \citep{2002AJ....123.2945R}. To address their underrepresentation in the training set derived from SDSS and DESI, we augment the stellar sample with the VLMS dataset compiled by \citet{2024ApJS..271...54F}, which includes 346{,}273 sources from a variety of surveys.

\subsubsection{The high-$z$ quasar sample}
We also compile a large high-redshift quasar sample to date, including 531 quasars with $z > 5.3$ from \citet{2023ARA&A..61..373F}, 450 quasars in the range $4.7 < z < 6.6$ from \citet{2023ApJS..269...27Y}, and 162 quasars from the SHELLQs project\footnote{\url{https://cosmos.phys.sci.ehime-u.ac.jp/~yk.matsuoka/shellqs.html}} \citep{2016ApJ...828...26M}. This sample is used exclusively to improve the photo-$z$ estimation by extending the training sets toward the high-redshift end. However, given its small fraction compared to the classification training set, its impact on classification performance is negligible, and thus it is not included in the classification process.

\begin{deluxetable}{lcccc}
\tablecaption{\centering Training and Test Set Sizes across Parent Samples. \label{tab:sample_number}}
\tablehead{
\colhead{Sample} & \colhead{$\mathrm{N_{Total}}$} & \colhead{$\mathrm{N_{QSO}}$} & \colhead{$\mathrm{N_{Galaxy}}$} & \colhead{$\mathrm{N_{Star}}$}
}
\startdata
\multicolumn{5}{c}{\textbf{Master Sample}} \\
\hline
Training Set & 3,078,645 & 241,181 & 2,458,060 & 379,404 \\
Random Test Set & 342,072 & 26,689 & 273,037 & 42,346 \\
HETDEX Test Set & 16,688 & 815 & 11,330 & 4,543 \\
VVDS Test Set & 12,031 & 162 & 11,869 & 0 \\
zCOSMOS Test Set & 16,302 & 225 & 16,077 & 0 \\
\hline
\multicolumn{5}{c}{\textbf{HSC+WISE Sample}} \\
\hline
Training Set & 1,655,637 & 171,060 & 1,275,056 & 209,521 \\
Random Test Set & 183,960 & 19,081 & 141,322 & 23,557 \\
HETDEX Test Set & 9,416 & 586 & 6,189 & 2,641 \\
VVDS Test Set & 3,181 & 101 & 3,080 & 0 \\
zCOSMOS Test Set & 5,968 & 175 & 5,793 & 0 \\
\hline
\multicolumn{5}{c}{\textbf{GoldenU Sample}} \\
\hline
Training Set & 1,194,169 & 68,302 & 949,687 & 176,180 \\
Random Test Set & 132,686 & 7,429 & 105,403 & 19,854 \\
HETDEX Test Set & 7,147 & 271 & 4,591 & 2,285 \\
VVDS Test Set & 2,897 & 90 & 2,807 & 0 \\
zCOSMOS Test Set & 4,144 & 86 & 4,058 & 0 \\
\hline
\multicolumn{5}{c}{\textbf{GoldenV Sample}} \\
\hline
Training Set & 1,090,178 & 87,296 & 855,434 & 147,448 \\
Random Test Set & 121,131 & 9,471 & 95,374 & 16,286 \\
HETDEX Test Set & 8,443 & 373 & 5,578 & 2,492 \\
VVDS Test Set & 2,240 & 71 & 2,169 & 0 \\
zCOSMOS Test Set & 5,783 & 168 & 5,615 & 0 \\
\enddata
\tablecomments{HETDEX, VVDS, and zCOSMOS sources are used exclusively for testing. SDSS, DESI, and VLMS sources are split into training and test sets with a 9:1 ratio. Duplicate entries in the training set are merged, and any sources present in test sets are removed from the training set.
}
\end{deluxetable}

\begin{figure*}[ht]
    \centering
    \includegraphics[width=1\textwidth]{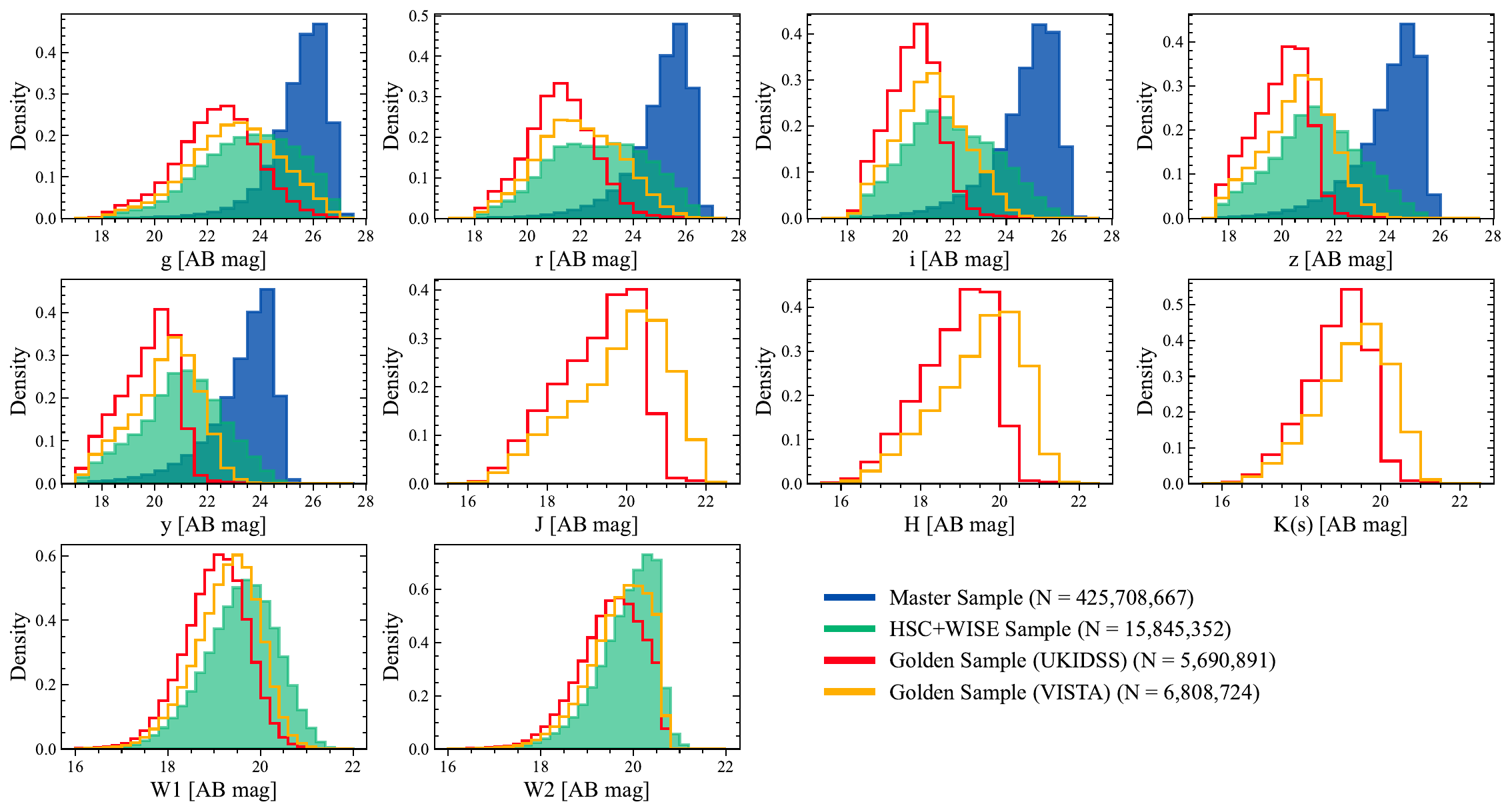}
    \caption{
    Histograms of apparent magnitudes for each parent samples in the HSC optical bands ($g$, $r$, $i$, $z$, $y$), UKIDSS near-IR bands ($J$, $H$, $K$), VISTA near-IR bands ($J$, $H$, $K_{\mathrm{s}}$), and WISE mid-IR bands ($W1$, $W2$). The y-axis represents the number density. The quantity of each parent sample is indicated in the legend.
    }
    \label{fig:sample_hist}
\end{figure*}

\subsection{Parent Samples} \label{subsec:parent_samples}

Considering the survey area and wavelength coverage, four parent samples have been constructed in this work. The first of these, the Master Sample, includes only HSC data and is designed to achieve the largest sample size. We select HSC sources that have at least one band with a valid aperture photometry measurement (i.e., $\texttt{phot\_flag} = 0$) and exhibit no image issues (i.e., $\texttt{image\_flag} = 0$) in at least three out of the five HSC bands. Such criteria maintain high data quality while retaining high-redshift dropout candidates. In addition, we incorporate mid-infrared photometry, which has been shown to be effective for quasar identification \citep{2005ApJ...631..163S,2019MNRAS.485.4539J}. All sources in the Master Sample are cross-matched with the CatWISE2020 catalog to obtain W1 and W2 photometry, yielding the HSC+WISE sample (hereafter \texttt{hsc\_wise}).

We incorporate near-IR photometry, which has been shown to significantly improve the accuracy of quasar photo-$z$ estimates \citep{2007MNRAS.375.1180C,2010MNRAS.406.1583W}. To extend near-IR coverage across more HSC fields, we cross-match sources in the HSC+WISE sample with the UKIDSS and VISTA surveys, resulting in two additional enhanced parent samples, hereafter referred to as \texttt{GoldenU} (HSC+WISE+UKIDSS) and \texttt{GoldenV} (HSC+WISE+VISTA). Figure~\ref{fig:sample_hist} presents the magnitude distributions across various bands for all four parent samples. Due to the relatively shallower depth of the WISE, UKIDSS, and VISTA surveys compared to the deep HSC survey, the HSC+WISE sample and the two golden samples are significantly brighter than the master sample. The sizes of the parent samples are as follows: 425,708,667 for the master sample, 15,845,352 for the HSC+WISE sample, 5,690,891 for the \texttt{GoldenU} sample, and 6,808,724 for the \texttt{GoldenV} sample.

\section{Machine-learning Selection of Quasar Candidates} \label{sec:ml_selection}

\begin{figure*}[ht]
    \centering
    \includegraphics[width=1\textwidth]{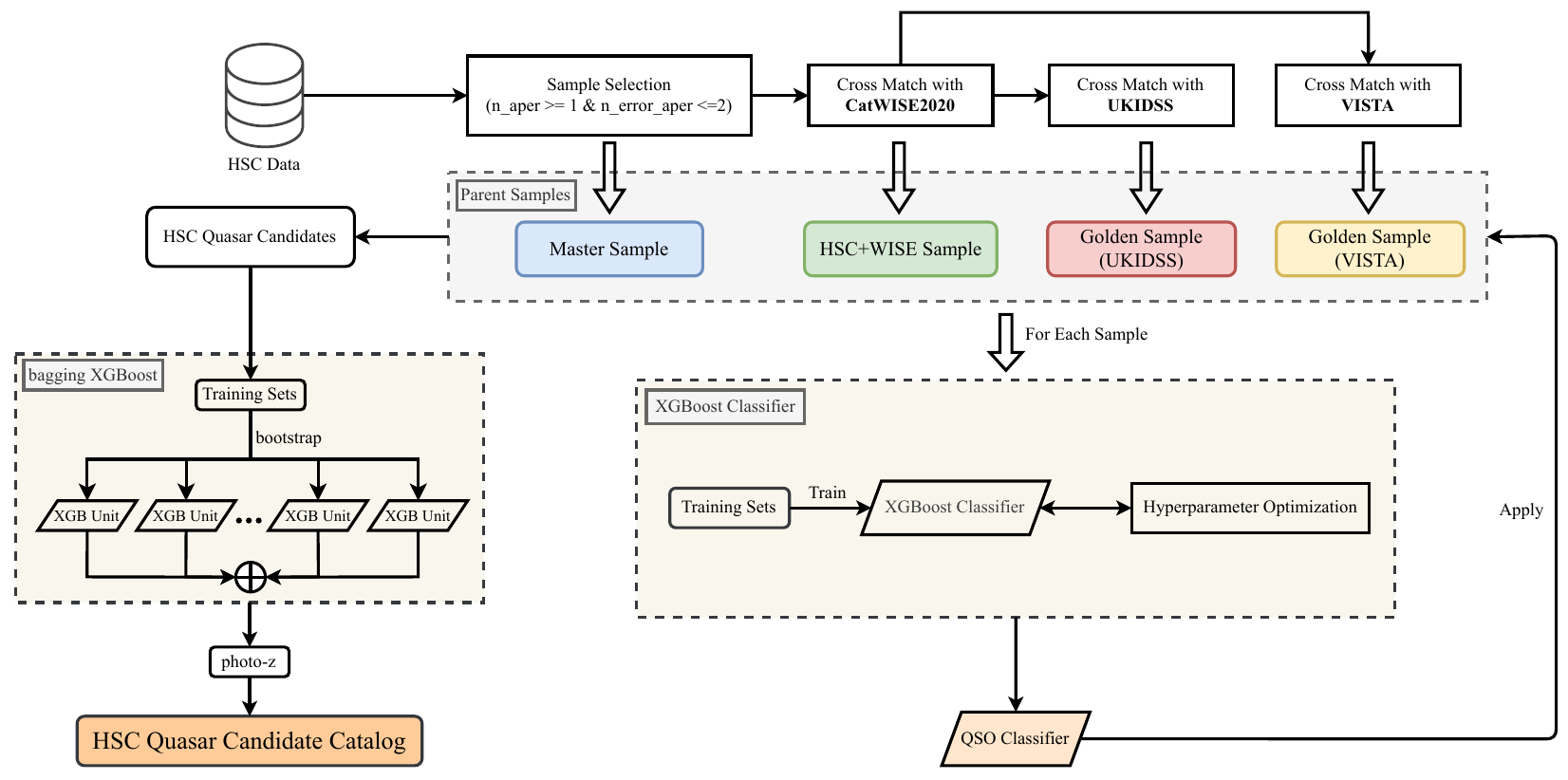}
    \caption{Simplified workflow of the QHSC pipeline. The pipeline starts by cross-matching the HSC data with inferred photometric catalogs (e.g., CatWISE2020, UKIDSS, and VISTA surveys) to build four parent samples. Each sample is used to train a separate XGBoost classifier, which selects quasar candidates and estimates their photo-$z$ using the bagging XGBoost method. The final output of the pipeline consists of the quasar candidate catalogs, along with their photo-$z$ estimates.}
    \label{fig:workflow}
\end{figure*}

\begin{deluxetable}{lcc}
\tablecaption{\centering Features Used to Train XGBoost Classifiers for Each Parent Sample \label{tab:features}}
\tablehead{
\colhead{Feature Class} & \colhead{Features} & \colhead{Samples}
}
\startdata
HSC Magnitudes & $(grizy)_{\mathrm{aper|psf|cmodel}}$ & All  \\
NIR Magnitudes  & $J$, $H$, $K$ & GoldenU  \\ 
NIR Magnitudes  & $J$, $H$, $K_{\mathrm{s}}$ & GoldenV  \\
MIR Magnitudes & $W1$, $W2$  & GoldenU, GoldenV, HSC+WISE \\
HSC Colors & $(g-r, r-i, i-z, z-y)_{\mathrm{aper|psf|cmodel}}$ & All \\
IR Colors & $y_{\mathrm{aper|psf|cmodel}}-J$, $J-H$, $H-K$, $K-W1$ & GoldenU  \\
IR Colors & $y_{\mathrm{aper|psf|cmodel}}-J$, $J-H$, $H-K_{\mathrm{s}}$, $K_{\mathrm{s}}-W1$ & GoldenV  \\
MIR Color & $W1-W2$ & GoldenU, GoldenV, HSC+WISE  \\
Extendedness & $X_{\mathrm{psf}} - X_{\mathrm{cmodel}}$, $X \in (grizy)$  & All
\enddata
\tablecomments{"All" refers to each of the parent samples: the master sample, HSC+WISE sample, GoldenU sample, and GoldenV sample.}
\end{deluxetable}

In this section, we apply machine learning techniques to select quasar candidates from each of the four parent samples using well-selected input features. Figure~\ref{fig:workflow} illustrates the entire process from the original HSC photometric catalog to the QHSC quasar candidate catalog, including photo-$z$ estimation using the bagging XGBoost regressor described in Section~\ref{sec:photoz}. Four independent XGBoost classifiers are trained separately, with hyperparameters optimized for each corresponding sample. Spectroscopic classifications are incorporated to construct the training and test sets. The final classifiers are then applied to their respective parent samples to identify quasar candidates, which are compiled into the QHSC catalog.

\subsection{Training Sets and Test Sets}

All spectroscopically classified sources (QSO, GALAXY, and STAR) with high confidence from the spectroscopic datasets (see Section~\ref{subsec:spec_samples}) are extracted and cross-matched with the parent samples using a $1''$ matching radius. Spectroscopically confirmed quasars, galaxies, and stars from the SDSS and DESI DR1 catalogs serve as the primary training datasets for the machine learning classification models. Additionally, stars from the VLMS sample are included as supplementary training data. 

\begin{figure*}[ht]
    \centering
    \includegraphics[width=1\textwidth]{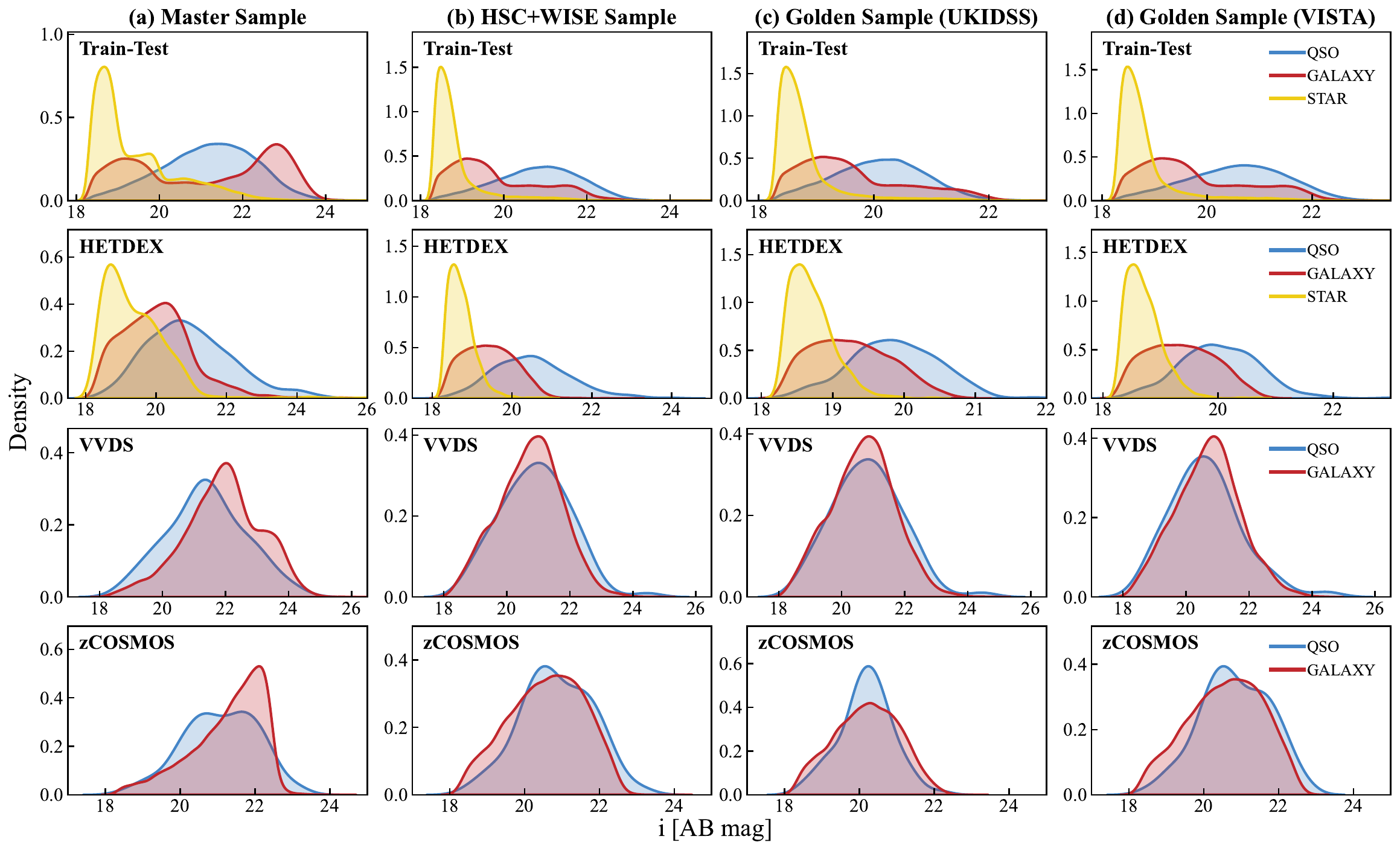}
    \caption{
    Distributions of $i$ band magnitude (number density) for the training and testing sets of each parent sample. The first row shows the combined SDSS, DESI, and VLMS samples used for model training and evaluation with a random 9:1 train test split. The remaining three rows, HETDEX, VVDS, and zCOSMOS, serve as independent test sets.
    }
    \label{fig:hist_train_test}
\end{figure*}

To evaluate classification performance, we construct four independent test sets using sources from the VVDS, zCOSMOS-bright, and HETDEX SC1 catalogs. All sources in these test sets are excluded from the input datasets (SDSS, DESI, and VLMS), and any objects with conflicting class labels are removed. Additionally, 10\% of the input data are randomly selected and reserved as a holdout test set (hereafter referred to as the random test set), while the remaining 90\% (the train+val set) are used for training and validation. These four test sets, drawn from surveys with varying spectroscopic strategies and depths, provide a comprehensive and robust assessment of classification performance. Table~\ref{tab:sample_number} summarizes the sizes of the training set and the various test sets for each parent sample, and Figure~\ref{fig:hist_train_test} shows the $i$-band magnitude distributions of the training and test sets. 
Despite the class imbalance among quasars, galaxies, and stars (1:10.2:1.6 for the master sample, 1:7.5:1.2 for HSC+WISE, 1:13.9:2.6 for GoldenU, and 1:9.8:1.7 for GoldenV), the sample sizes of all classes are sufficiently large to ensure statistical robustness and to support the training of effective classifiers.

\subsection{Features} \label{subsec:classifier_features}
Colors derived from identical photometric types (e.g., \texttt{CModel}, \texttt{PSF}, aperture) serve as crucial features for both quasar classification and photometric redshift (photo-$z$) estimation \citep[see e.g.,][]{2019MNRAS.485.4539J,2024ApJS..271...54F}. In this study, we utilize three photometric measurements from the HSC catalogs: (1) $2''$ aperture photometry (\texttt{aper}), (2) \texttt{PSF} photometry, and (3) \texttt{CModel} photometry. Each photometric type offers distinct advantages: \texttt{PSF} photometry provides optimal measurements for point sources, while \texttt{CModel} performs better for extended sources. The aperture photometry, being model-independent, offers robust measurements that are less sensitive to model assumptions.  
Given the importance of morphological information for classification \citep{2024ApJS..271...54F}, we include the magnitude differences between \texttt{PSF} and \texttt{CModel} measurements in each HSC band as a key morphological feature to characterize the extendedness. We deliberately omit extinction corrections for magnitudes during classification, as these corrections require prior knowledge of source types (whether Galactic or extragalactic) and distances (for Galactic objects), which is unavailable before the classification.  

To ensure the robustness of our feature selection, we conducted a validation process by introducing a random noise feature into the feature set. This allowed us to verify that all selected features exhibited higher importance scores than the noise feature during training. The noise feature was subsequently excluded from the final model. Table~\ref{tab:features} summarizes all features used in our analysis. The total feature counts are 44 for the Golden Sample (UKIDSS), 44 for the Golden Sample (VISTA), 38 for the HSC+WISE Sample, and 32 for the Master Sample.

\subsection{XGBoost Classifier Model Training}\label{subsec:classification}

We adopt XGBoost \citep{2016arXiv160302754C}, an efficient implementation of the gradient boosting decision tree algorithm, to train classification models for each parent sample. The XGBoost objective function comprises a training loss term, which evaluates predictive accuracy, and a regularization term, which controls model complexity. We use the multiclass log loss (i.e., categorical cross-entropy) as the optimization objective, defined as:
\begin{gather}
    L(y, \hat{p}) = -\sum_{i=1}^C y_i \ln(\hat{p}_i) \,,
\end{gather}
where $y_i$ denotes the true label from the one-hot encoded target vector, $\hat{p}_i$ is the predicted probability for class $i$ produced by the softmax function, and $C$ is the number of classes. The regularization term includes several hyperparameters to penalize model complexity and prevent overfitting. 

The input features are used without any imputation for missing values, as the missing rate in the parent samples is low and XGBoost inherently supports handling missing data during the training. In view of the class imbalance among quasars, stars, and galaxies in the training set, we explored several mitigation strategies, including data resampling. However, these approaches did not lead to significant performance improvements. As a result, no specific imbalance correction techniques are applied in this study.

A validation set is created by randomly splitting the train+val dataset with a 4:1 train-to-validation ratio. Hyperparameter optimization is performed using \texttt{Optuna} \citep{2019arXiv190710902A}, aiming to minimize the validation log loss. The number of boosting rounds (\texttt{num\_boost\_round}) is fixed at 700, 250, 200, and 200 for the master, HSC+WISE, GoldenU, and GoldenV samples, respectively, reflecting the varying training set sizes and complexity. For each parent sample, 500 \texttt{Optuna} trials are conducted on an NVIDIA A100 GPU. Final models are trained on the full train+val datasets using the optimized hyperparameters. Table~\ref{tab:optuna} summarizes the default and optimized hyperparameters along with corresponding performance metrics.

\subsection{Performance Metrics for Classification}

\begin{figure*}[ht]
    \centering
    \includegraphics[width=1\textwidth]{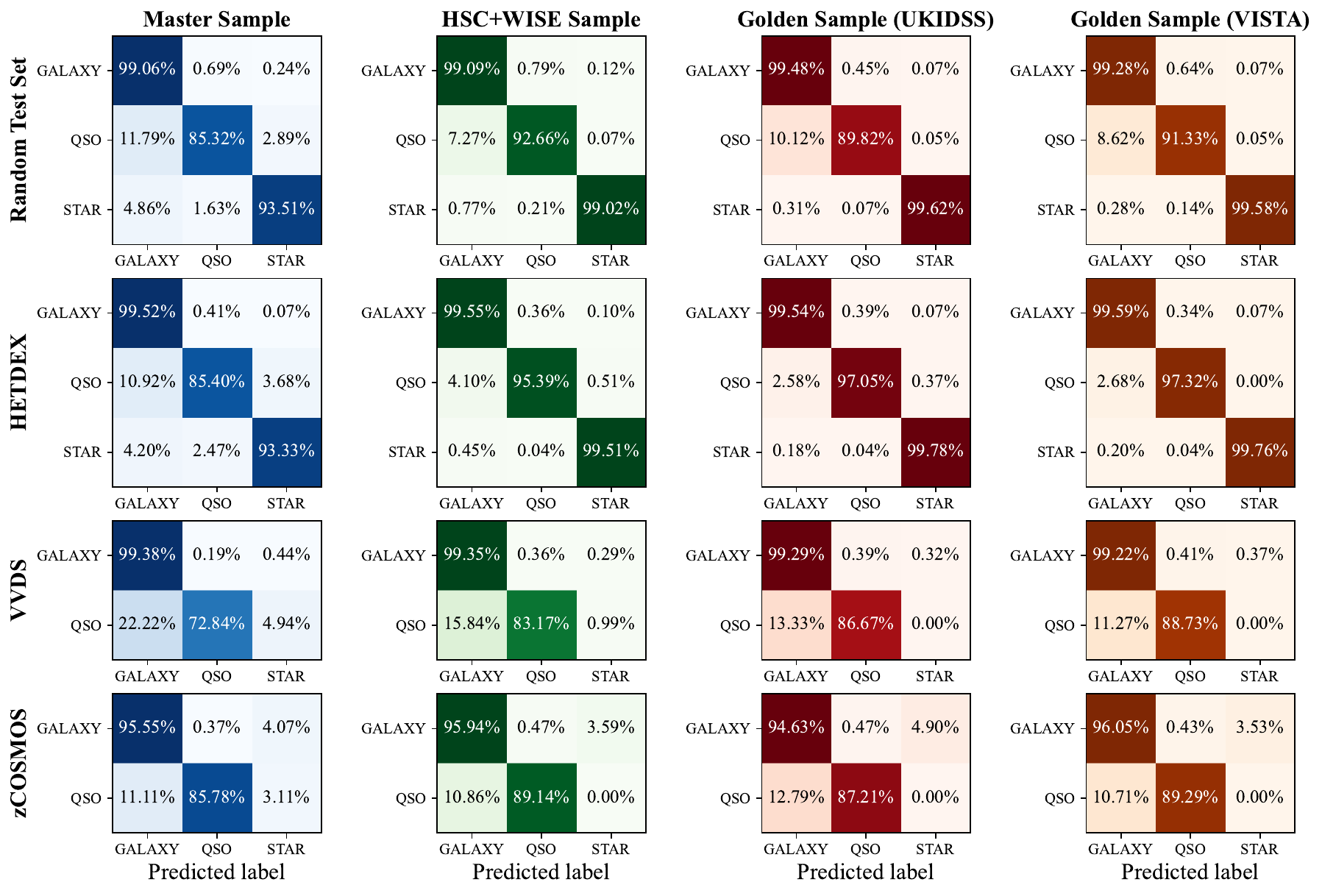}
    \caption{
    Normalized confusion matrices of XGBoost classifiers for each parent sample, evaluated on the random, HETDEX, VVDS, and zCOSMOS test sets. The color scale reflects the number of sources. Diagonal elements denote recall (completeness) for each class and off-diagonal elements represent misclassification rates.
    }
    \label{fig:confusion_matrix}
\end{figure*}

To evaluate the performance of the classifiers, we adopt four commonly used metrics: precision, recall, area under the precision-recall curve (AUCPR), and the weighted F1 score. Precision, recall, and AUCPR are specifically used to assess the classification performance for the QSO class, while the weighted F1 score provides an overall evaluation that accounts for class imbalance. All metrics are computed using the \texttt{sklearn.metrics} module from the scikit-learn package \citep{2011JMLR...12.2825P}.

For the QSO-specific metrics, the classification task is treated as a binary problem, where the QSO class is considered positive, and the GALAXY and STAR classes are grouped as negative. The definitions of the metrics based on true positives (TP), false positives (FP), false negatives (FN), and true negatives (TN) are as follows:
\begin{equation}
    \text{Precision} = \frac{\text{TP}}{\text{TP} + \text{FP}} \,,
\end{equation}

\begin{equation}
    \text{Recall} = \frac{\text{TP}}{\text{TP} + \text{FN}} \,,
\end{equation}

\begin{equation}
    \text{F1} = \frac{2 \times \text{Precision} \times \text{Recall}}{\text{Precision} + \text{Recall}} \,,
\end{equation}

\begin{equation}
    \text{Weighted F1} = \frac{\sum_{k=1}^{K} N_k \cdot \text{F1}(k)}{\sum_{k=1}^{K} N_k} \,,
\end{equation}
where $N_k$ denotes the number of samples in class $k$, and $F1(k)$ is the F1 score for that class.

\begin{figure*}[ht]
    \centering
    \includegraphics[width=1\textwidth]{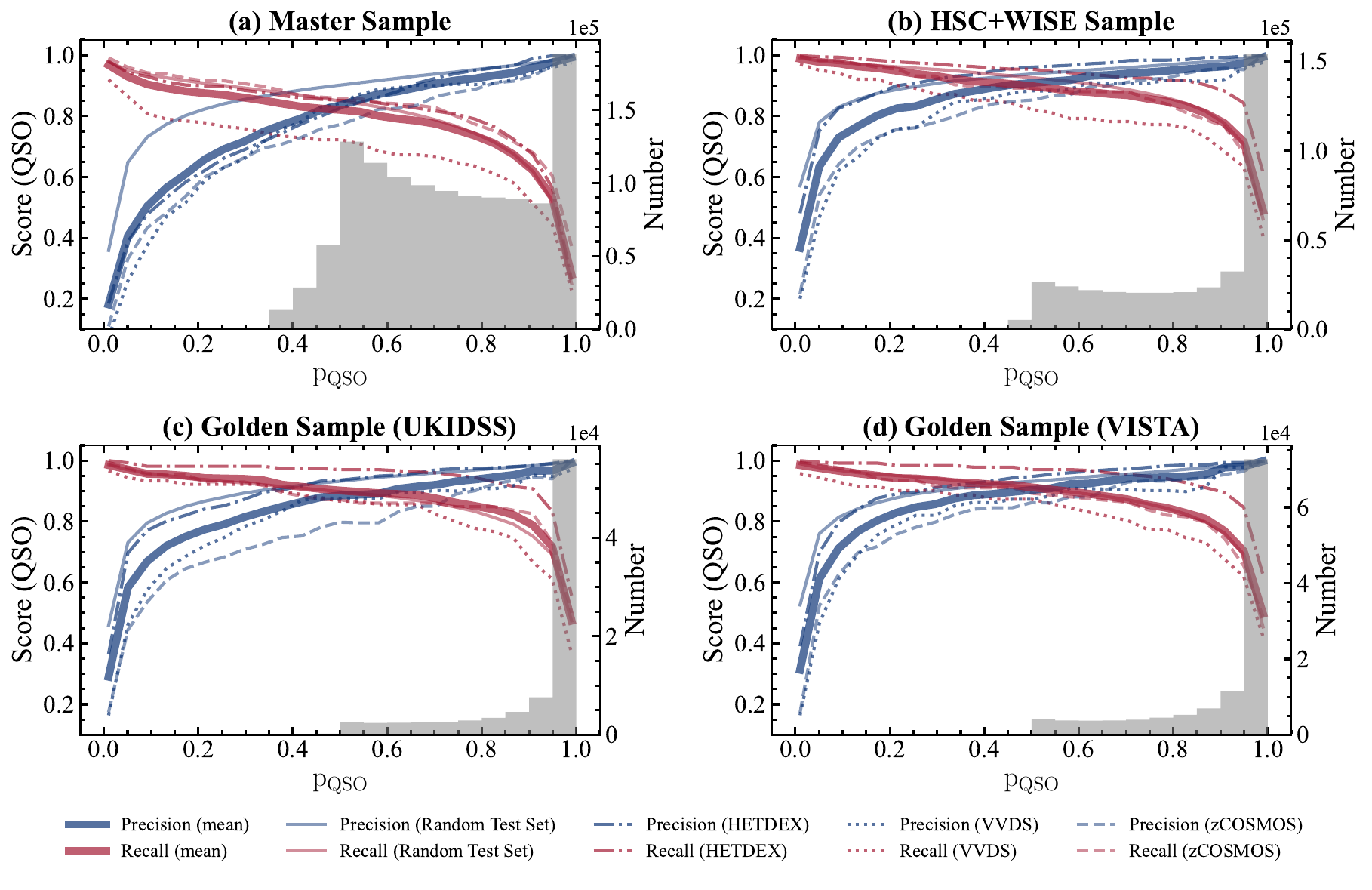}
    \caption{
    Precision (blue) and recall (red) as functions of the quasar classification probability ($\pqso$) threshold, indicated on the left y-axis. Different line styles represent distinct test sets, with the thick lines denoting the mean values across all four test sets. The right y-axis displays the histogram of $\pqso$ values for all QHSC candidates.
    }
    \label{fig:pr_threshold}
\end{figure*}

\begin{deluxetable}{lcccccccc}
\tablecaption{Comparison of XGBoost Classifier Performance Using Default versus Optimized Hyperparameter Settings. \label{tab:optuna}} 
\tabletypesize{\footnotesize}
\tablehead{
\colhead{Hyperparameter/Metrics} & \multicolumn{2}{c}{Master Sample} & \multicolumn{2}{c}{HSC+WISE Sample} & \multicolumn{2}{c}{Golden Sample (UKIDSS)} & \multicolumn{2}{c}{Golden Sample (VISTA)}\\
\cline{2-9}
\colhead{} & \colhead{Default} & \colhead{Optimal} & \colhead{Default} & \colhead{Optimal} & \colhead{Default} & \colhead{Optimal} & \colhead{Default} & \colhead{Optimal}
} 
\startdata
\texttt{num\_boost\_round} & 10 & 700 & 10 & 250 & 10 & 200 & 10 & 200 \\
\texttt{max\_depth} & 6 & 12 & 6 & 12 & 6 & 11 & 6 & 11 \\
\texttt{eta} & 0.3 & 0.08 & 0.3 & 0.15 & 0.3 & 0.14 & 0.3 & 0.14 \\
\texttt{alpha} & 0 & 7.52 & 0 & 9.28 & 0 & 9.60 & 0 & 7.36 \\
\texttt{lambda} & 1 & 1.43 & 1 & 0.61 & 1 & 1.87 & 1 & 1.08 \\
\texttt{gamma} & 0 & 0.00 & 0 & 0.19 & 0 & 0.21 & 0 & 0.08 \\
\texttt{min\_child\_weight} & 1 & 3 & 1 & 5 & 1 & 1 & 1 & 5 \\
\texttt{subsample} & 1 & 0.76 & 1 & 0.88 & 1 & 0.82 & 1 & 0.91 \\
\texttt{colsample\_bytree} & 1 & 0.71 & 1 & 0.53 & 1 & 0.52 & 1 & 0.51 \\
\texttt{max\_delta\_step} & 0 & 1.61 & 0 & 3.18 & 0 & 4.35 & 0 & 2.68 \\
Weighted F1 & 0.9635 & 0.9728 & 0.9804 & 0.9841 & 0.9875 & 0.9895 & 0.9840 & 0.9870 \\
AUCPR (QSO) & 0.8723 & 0.9349 & 0.9648 & 0.9782 & 0.9490 & 0.9666 & 0.9563 & 0.9725 \\
Precision (QSO) & 0.8413 & 0.8983 & 0.9263 & 0.9378 & 0.9303 & 0.9319 & 0.9232 & 0.9317 \\
Recall (QSO) & 0.7885 & 0.8532 & 0.9072 & 0.9266 & 0.8659 & 0.8982 & 0.8849 & 0.9133 \\
\enddata
\end{deluxetable}

\subsection{Evaluation of Classifiers} \label{subsec:classifiers_evaluation}

The normalized confusion matrices computed for four test sets, shown in Figure~\ref{fig:confusion_matrix}, provide a detailed assessment of the  performance of the optimal XGBoost classifiers. Using only five HSC broadband filters, approximately 85\% of quasars are correctly identified in the random test set. When WISE data are incorporated, the recall rate increases to about 90\%. The other three test sets exhibit varying performance levels due to the differences in sample depth. A more detailed discussion of the performance across different magnitude bins is presented in Section~\ref{subsec:bin_evaluation}.

The outputs of the XGBoost classifiers are the predicted probabilities for each source to be a quasar, galaxy, or star, denoted as $\pqso$, $p_{\mathrm{GAL}}$, and $p_{\mathrm{STAR}}$, respectively. Quasar candidates are defined as sources whose quasar probability $\pqso$ exceeds both $p_{\mathrm{GAL}}$ and $p_{\mathrm{STAR}}$. Finally, the numbers of quasar candidates are 87,460 (goldenU), 120,572 (goldenV), 371,777 (HSC+WISE), and 1,184,574 (Master). Additionally, $\pqso$ quantifies the similarity between a source and the quasars in the training set in feature space. A purer quasar subsample can be obtained by imposing a threshold on $\pqso$. The variation of recall (completeness) and precision (purity) with respect to the $\pqso$ threshold is shown in Figure~\ref{fig:pr_threshold}. By selecting an appropriate threshold, one can obtain either a purer or a more complete quasar candidate subsample. 

\begin{figure*}[ht]
    \centering
    \includegraphics[width=1\textwidth]{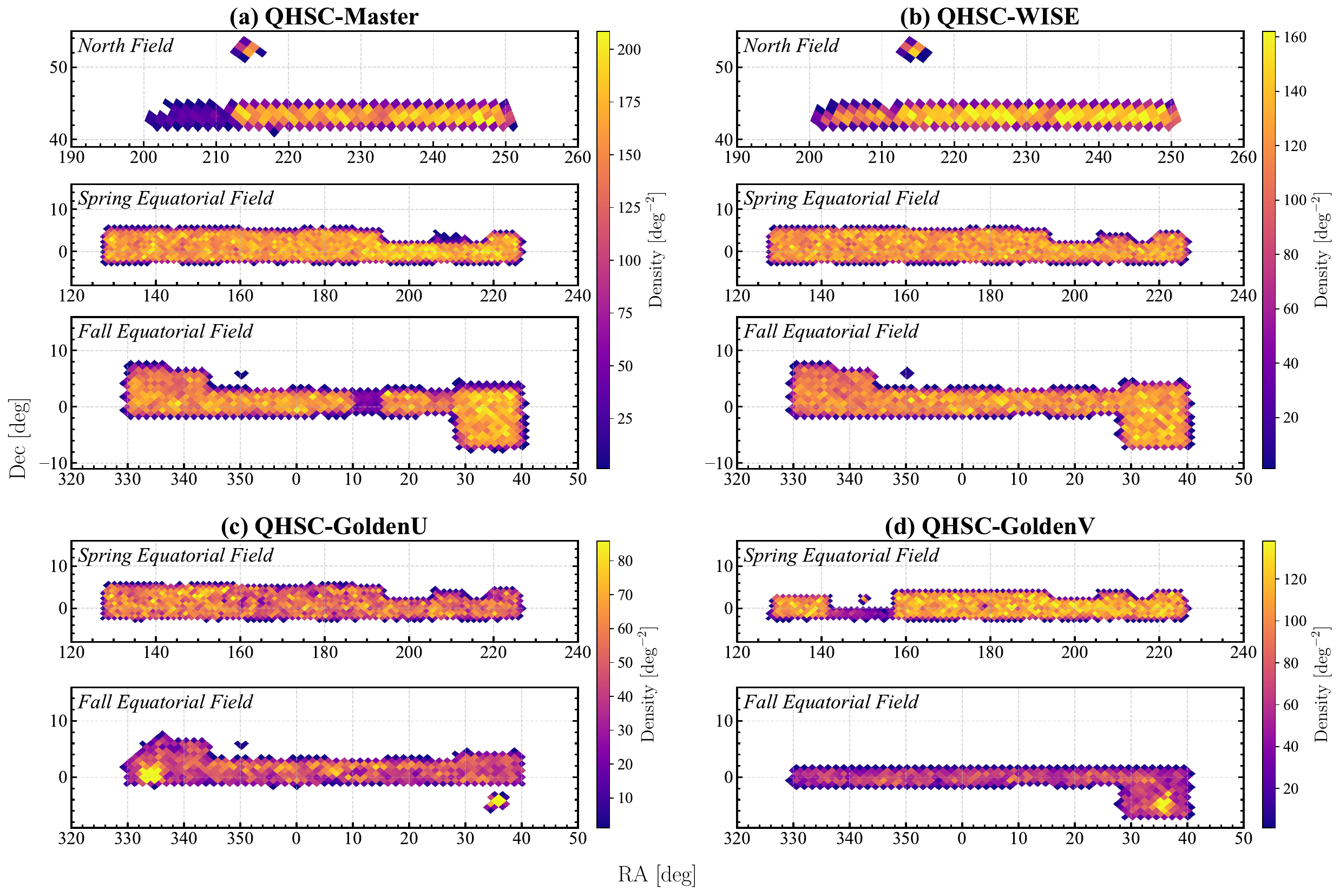}
    \caption{
    HEALPix number density maps of QHSC quasar candidates (high-purity subsample with a threshold of $p_{\mathrm{QSO}} > 0.95$). Each panel corresponds to an individual sub-field of the HSC Wide Survey. The maps are shown in the Equatorial coordinate system with a resolution parameter of $N_{\mathrm{side}} = 64$, corresponding to an area of 0.839\,deg$^2$ per pixel.
    }
    \label{fig:number_density}
\end{figure*}

Figure~\ref{fig:number_density} presents the sky density maps of the high-purity subsample ($\pqso > 0.95$) of QHSC candidates. A prominent high-density region around $\mathrm{RA} \sim 334^\circ$ is observed in the HSC Fall Equatorial Field of the GoldenU sample. This enhancement is attributed to the deeper near-IR photometry provided by the UKIDSS-DXS in the SA22 field \citep{2007MNRAS.379.1599L}, centered at $\mathrm{RA} = 22^{\mathrm{h}}17^{\mathrm{m}},\ \mathrm{Dec} = +00^{\circ}20^{\prime}$, which also overlaps with the VVDS-22h 2217+00 field \citep{2013A&A...559A..14L}. Another similar high-density region appears at $\mathrm{RA} \sim 36^\circ$ in the same HSC field, visible in both the GoldenU and GoldenV samples. This is primarily due to the presence of deeper near-IR data from the UKIDSS-UDS and the VISTA VIDEO-XMM field, centered at $\mathrm{RA} = 2^{\mathrm{h}}25^{\mathrm{m}},\ \mathrm{Dec} = -04^{\circ}30^{\prime}$. The median densities of the high-purity quasar candidates are $147.74\ \mathrm{deg}^{-2}$ (master), $114.38\ \mathrm{deg}^{-2}$ (HSC+WISE), $47.66\ \mathrm{deg}^{-2}$ (GoldenU), and $73.87\ \mathrm{deg}^{-2}$ (GoldenV).

\begin{figure*}[ht]
    \centering
    \includegraphics[width=1\textwidth]{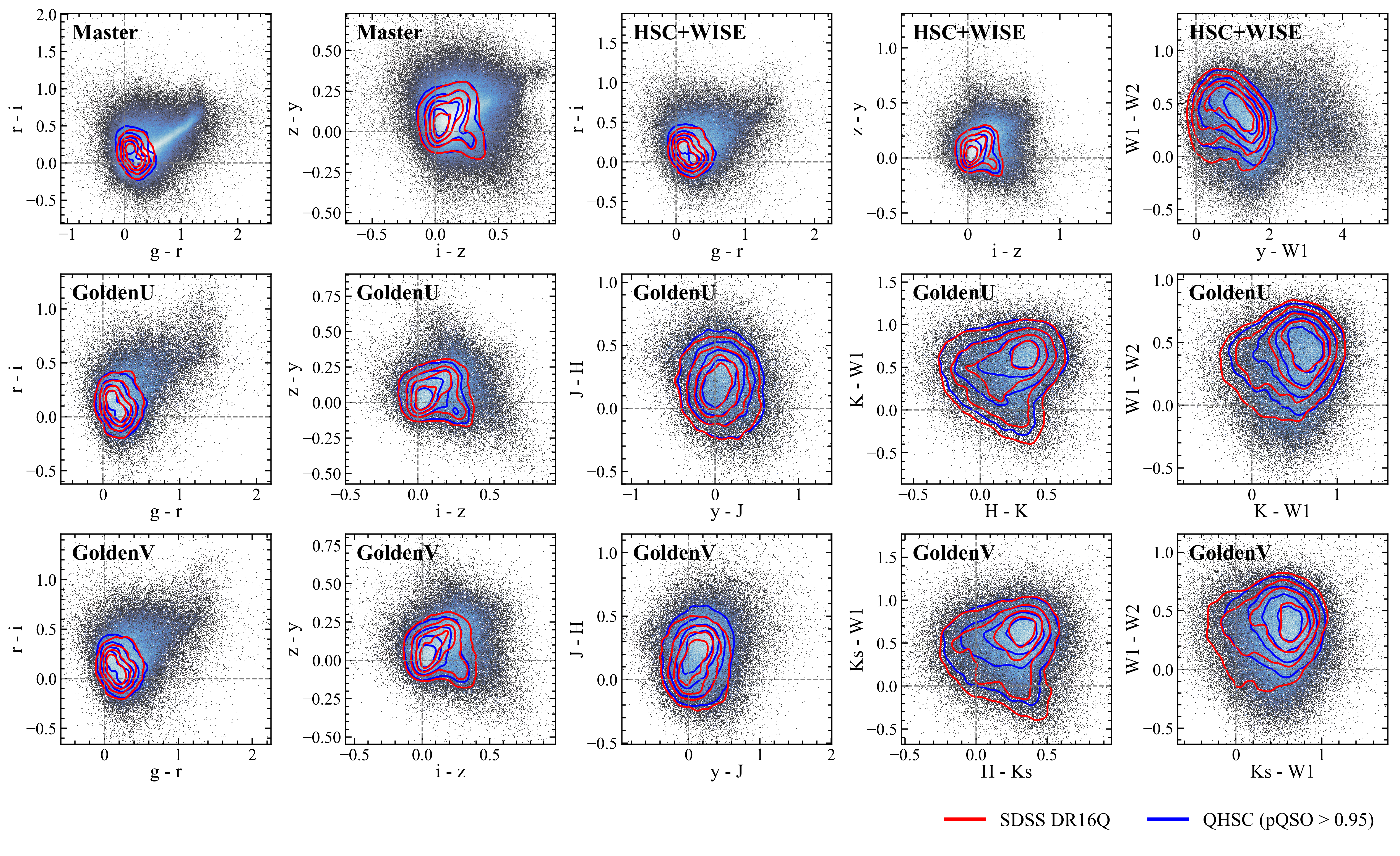}
    \caption{
    Color-color diagrams of sources from the QHSC quasar candidate catalog (blue density plots), the purified QHSC subsample with a $\pqso$ threshold of 0.95 (blue contours), and SDSS DR16Q quasars for each parent sample. HSC magnitudes are based on CModel photometry, and all magnitudes are in the AB system.
    }
    \label{fig:QHSC_ccplot}
\end{figure*}

The color--color distributions of QHSC and SDSS DR16Q quasars are compared in Figure~\ref{fig:QHSC_ccplot}. For the Master sample, a prominent stellar locus (narrow stripe) appears in the $gri$ color--color diagram, indicating considerable stellar contamination in the classification results. This issue can be alleviated by applying a threshold on the quasar probability $\pqso$. When a threshold of $\pqso > 0.95$ is adopted, the purified QHSC subsample aligns well with the DR16Q quasar distribution.

\begin{figure*}[ht]
    \centering
    \includegraphics[width=1\textwidth]{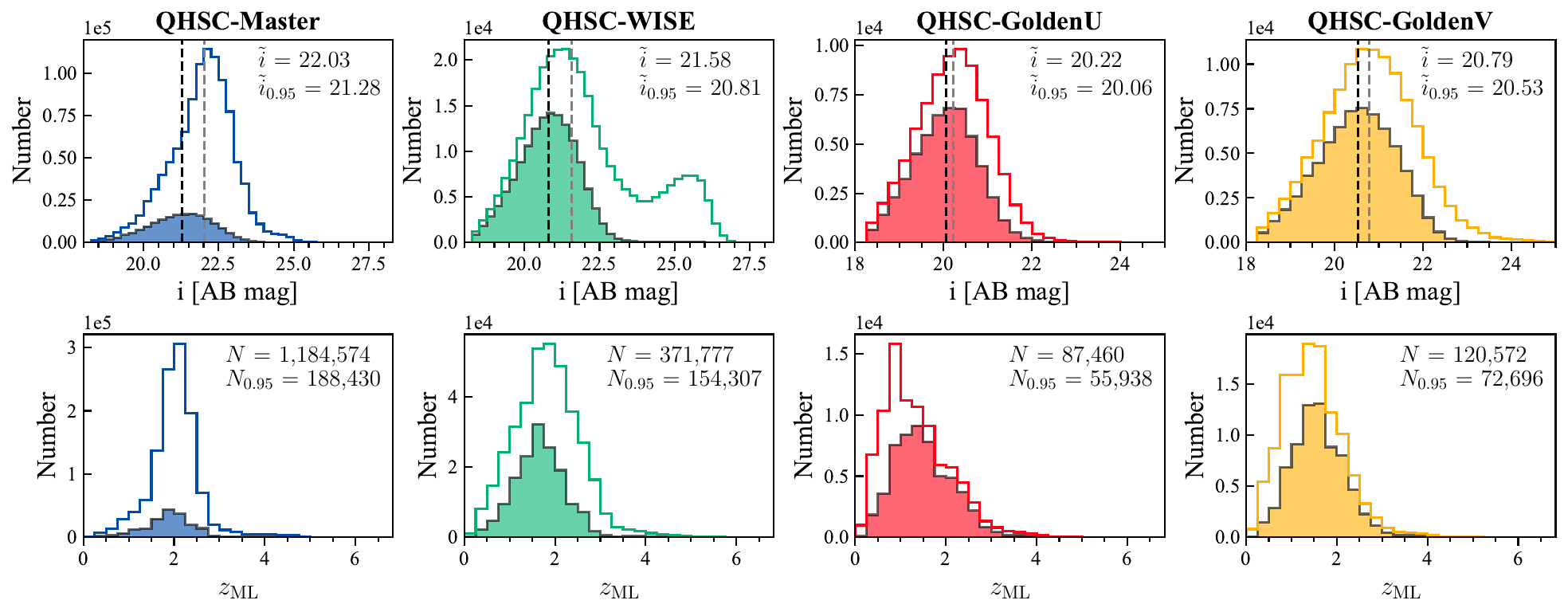}
    \caption{
    Distributions of HSC $i$-band (CModel) magnitudes and photometric redshifts for all quasar candidates (unfilled) and the high-purity subsample (filled) with $\pqso > 0.95$ from the QHSC catalog. The median $i$-band magnitudes for all candidates ($\tilde{i}$) and for the high-purity subsample ($\tilde{i}_{0.95}$) are indicated in the legends of the upper panels. The total numbers of objects ($N$) and of the high-purity subsample ($N_{0.95}$) are indicated in the legends of the lower panels.
    }
    \label{fig:QHSC_mag_z_hist}
\end{figure*}

The inclusion of WISE data further reduces stellar contamination, since mid-infrared colors exploit the stronger infrared excess of quasars relative to stars, arising from the power-law emission of the central engine and the thermal emission of surrounding dust. In the HSC+WISE sample, many faint sources are selected due to their red $\text{y} - \text{W1}$ colors (see Figure~\ref{fig:QHSC_mag_z_hist}). However, this red population is unlikely to be reliable quasar candidates, as most of these sources are very faint ($i > 24$), falling outside the magnitude range covered by the training set.

\subsection{Binned Evaluation Using Deep Spectroscopic Surveys} \label{subsec:bin_evaluation}

\begin{figure*}[ht]
    \centering
    \includegraphics[width=1\textwidth]{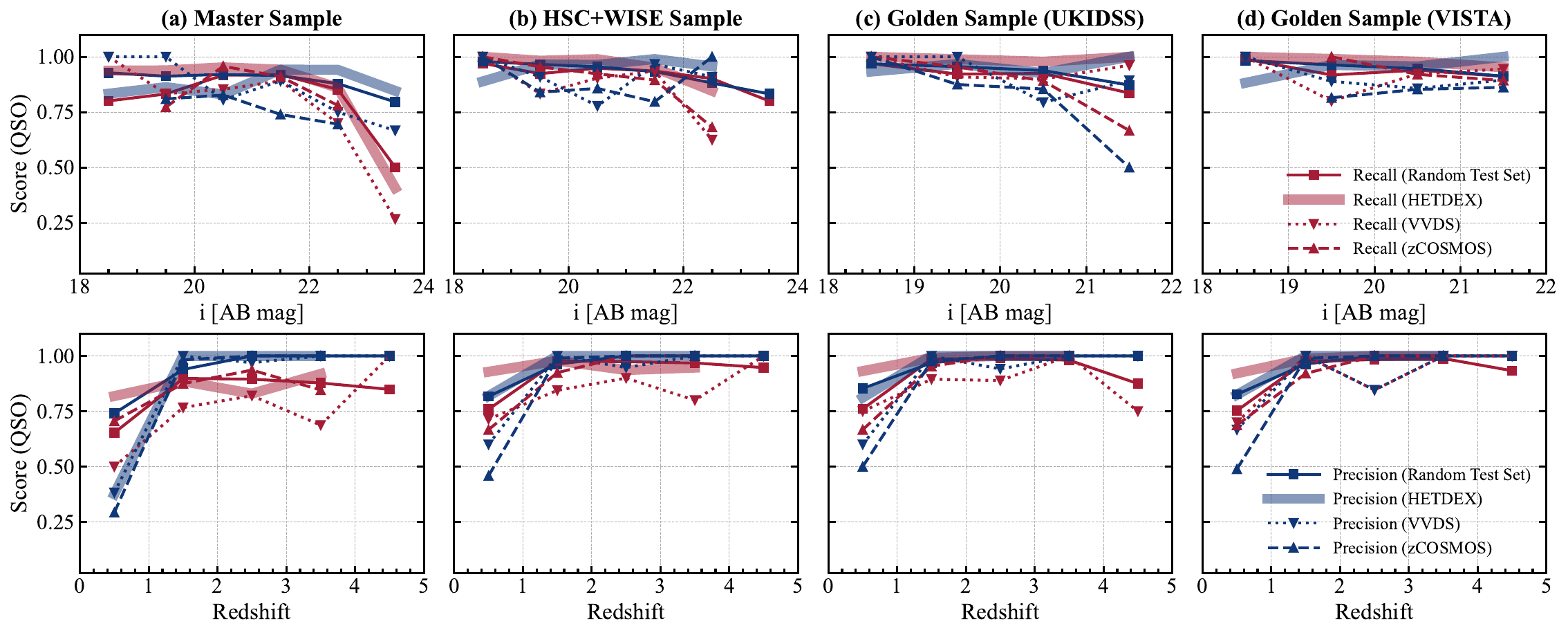}
    \caption{
    The recall (completeness, red) and precision (purity, blue) of quasar selection by XGBoost classifiers across different apparent magnitudes (top panels) and redshifts (bottom panels) for four parent samples. The results are based on four test sets: random test (solid line with square marker), HETDEX (broad line), VVDS (dotted line with downward triangle marker), and zCOSMOS-bright (dashed line with upward triangle marker). Results are omitted for bins with fewer than 3 test quasars or predicted quasars.
    }
    \label{fig:PR_bins}
\end{figure*}

\begin{figure*}[ht]
    \centering
    \includegraphics[width=1\textwidth]{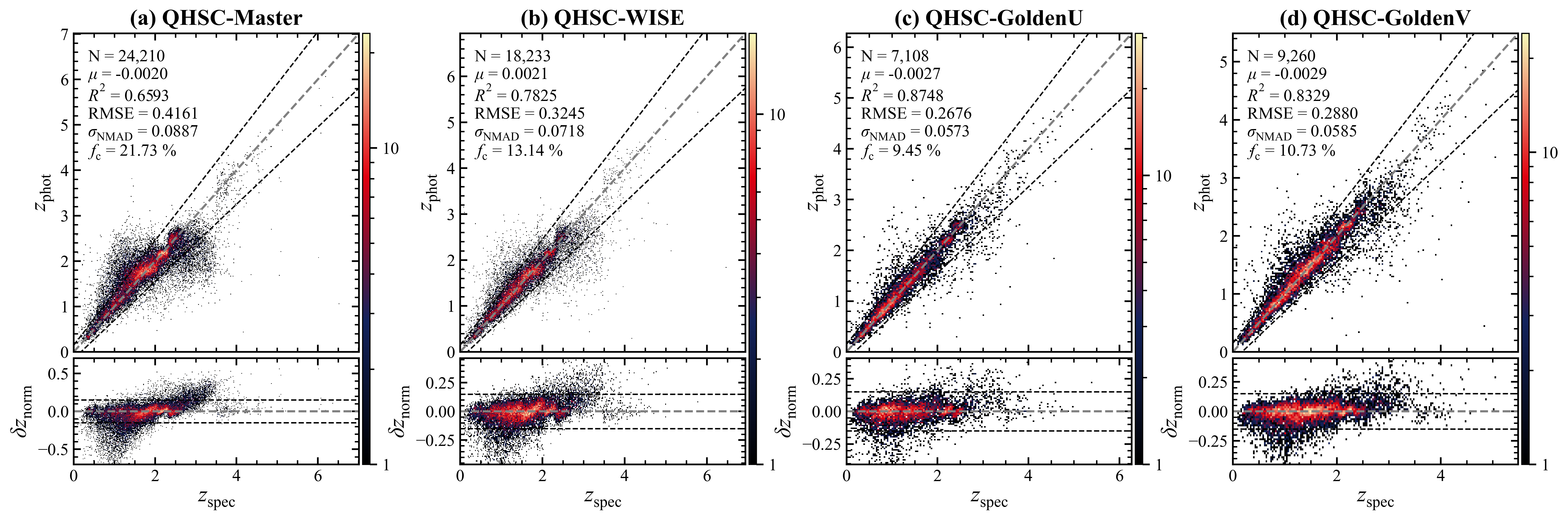}
    \caption{
    Photometric redshifts ($z_{\mathrm{phot}}$) predicted by the bagging XGBoost models (see Section~\ref{subsec:bsXGBoost}) compared to the spectroscopic redshifts ($z_{\mathrm{spec}}$). The comparison is based on the test sets, which are randomly split from the input datasets with a 9:1 train-test ratio. The top panels show the 2D density distributions of the predicted versus true redshifts, where the color scale indicates the number of sources per pixel. The bottom panels display the normalized redshift residuals, defined as $\delta z_{\mathrm{norm}} = (z_{\mathrm{spec}} - z_{\mathrm{phot}})/(1 + z_{\mathrm{spec}})$, as a function of $z_{\mathrm{spec}}$. The gray dashed lines mark the 1:1 relation ($z_{\mathrm{phot}} = z_{\mathrm{spec}}$), while the black dashed lines indicate the boundaries of the catastrophic outlier region at $z_{\mathrm{phot}} = z_{\mathrm{spec}} \pm 0.15(1 + z_{\mathrm{spec}})$.
    }
    \label{fig:photoz}
\end{figure*}

The HETDEX, VVDS, and zCOSMOS-bright surveys provide independent test sets to evaluate the performance of our quasar selection classifiers. The classifiers' performance varies across different brightness and redshift ranges. In Figure~\ref{fig:PR_bins}, we present the precision and recall rates of quasars across different magnitude and redshift bins. Results from all four test sets demonstrate that the classifiers maintain consistently high precision at redshifts $z > 1$. The observed decline in both precision and recall at $z < 1$ likely results from the under-representation of low-redshift quasars in our training set from SDSS and DESI spectroscopic surveys. For the master sample, Figure~\ref{fig:PR_bins} (a) reveals a lower recall rate at the faint end ($i > 23$), which is attributed to the depth limitation of the spectroscopic training set, making it challenging for the model to perform well on the faintest sources.

The precision and recall scores tested with the random test set are consistently higher than those from the other test sets, as the random test set closely mirrors the training sets. However, the other three test sets better represent real-world scenarios, encompassing a wider range of conditions, such as varying source brightness and redshift distributions, and provide an indicative assessment of the model's generalizability.

The reduced performance at the faint end ($i > 23$), particularly in the master sample, reflects the limited generalization capability of the machine learning model, arising from the restricted luminosity coverage of the available spectroscopic training sets. Enhancing classification reliability in this regime will require the inclusion of deeper spectroscopic samples that probe fainter magnitudes. Upcoming large-scale surveys such as the Subaru Prime Focus Spectrograph \citep[PFS;][]{2014PASJ...66R...1T} are expected to substantially augment the faint-end spectroscopic coverage, thereby improving the model's ability to identify quasars at the faint end.

\section{Photometric Redshift Estimation for Quasar Candidates} \label{sec:photoz}

\subsection{Ensemble Learning Using XGBoost as the Base Learner} \label{subsec:bsXGBoost}
Accurate redshift estimation is essential for leveraging the quasar candidate catalog in a broad range of applications, including follow-up studies and precision cosmology. Multi-band photometry provides a sparse sampling of the SED, which can be sufficient for estimating quasar photo-$z$ by capturing their broad spectral features, including the continuum shape and prominent emission and absorption lines \citep{2019NatAs...3..212S}. The photo-$z$ estimation task can be effectively framed as a regression problem within the machine learning paradigm.

All reliable quasar redshifts from the spectroscopic surveys described in Section~\ref{subsec:spec_samples} are used to build the photo-$z$ regression models. After removing duplicate entries, the fractions of quasars with known spectroscopic redshifts (spec-$z$) in the QHSC catalog are $81.47\%$, $76.36\%$, $48.83\%$, and $20.32\%$ for the goldenU, goldenV, HSC+WISE, and master samples, respectively. The sufficiently large number of known spec-$z$ quasars in the four catalogs enables both the training of machine learning regression models and the evaluation of their performance.

Ensemble learning methods are known to effectively enhance model performance. The XGBoost algorithm is an ensemble of decision trees trained via a boosting strategy. Although a single XGBoost regressor already provides high accuracy and efficiency, we further improve the robustness and predictive accuracy by adopting a bootstrap aggregating (bagging) approach, using XGBoost as the base learner. The feature set employed in photo-$z$ estimation is identical to that used for classification, as shown in Table~\ref{tab:features}.

As illustrated in Figure~\ref{fig:workflow}, the ensemble model (bagging XGBoost regressor) first generates 100 bootstrap resampled training sets, each used to train an independent XGBoost regressor. The final predicted redshift (\texttt{z\_ML}) is computed as the mean of the outputs from all base learners. Additionally, the standard deviation of the predictions (\texttt{z\_ML\_std}) serves as an estimate of the model uncertainty for each predicted photo-$z$.

\begin{deluxetable}{lcccccccccccc}
\tablecaption{Assessing Quality of Photo-$z$ for the QHSC catalogs. \label{tab:photoz}} 
\tabletypesize{\scriptsize}
\tablehead{
\colhead{Metrics} & \multicolumn{3}{c}{Master Sample} & \multicolumn{3}{c}{HSC+WISE Sample} & \multicolumn{3}{c}{Golden Sample (UKIDSS)} & \multicolumn{3}{c}{Golden Sample (VISTA)}\\
\hline
\colhead{} & \colhead{Single} & \colhead{Bagging} & \colhead{STD Cut} & \colhead{Single} & \colhead{Bagging} & \colhead{STD Cut} & \colhead{Single} & \colhead{Bagging} & \colhead{STD Cut} & \colhead{Single} & \colhead{Bagging} & \colhead{STD Cut}
} 
\startdata
$\mu$ & -0.0023 & -0.0020 & -0.0010 & 0.0024 & 0.0021 & 0.0003 & -0.0050 & -0.0027 & -0.0005 & -0.0034 & -0.0029 & -0.0026 \\
$R^2$ & 0.6462 & 0.6593 & 0.7325 & 0.7706 & 0.7825 & 0.8361 & 0.8595 & 0.8748 & 0.9353 & 0.8198 & 0.8329 & 0.9017 \\
RMSE & 0.4241 & 0.4161 & 0.3236 & 0.3332 & 0.3245 & 0.2377 & 0.2836 & 0.2676 & 0.1527 & 0.2990 & 0.2880 & 0.1754 \\
$\sigma_{\mathrm{NMAD}}$ & 0.0937 & 0.0887 & 0.0680 & 0.0756 & 0.0718 & 0.0613 & 0.0655 & 0.0573 & 0.0466 & 0.0667 & 0.0585 & 0.0463 \\
$f_c$ & 0.2250 & 0.2173 & 0.1276 & 0.1402 & 0.1314 & 0.0769 & 0.1086 & 0.0945 & 0.0367 & 0.1188 & 0.1073 & 0.0445 \\
\hline
N & 24,210 & 24,210 & 16,574 & 18,233 & 18,233 & 13,699 & 7,108 & 7,108 & 4,767 & 9,260 & 9,260 & 6,198 \\
\enddata
\tablecomments{
The “Single” columns report the performance of the single XGBoost regressor on the test sets. The “Bagging” columns show the results obtained using the bagging XGBoost ensemble models. The “STD Cut” columns present the performance of the bagging XGBoost models evaluated on subsamples of the test sets with lower model uncertainty ($\texttt{z\_ML\_std} < 0.1$).
}
\end{deluxetable}

\begin{figure*}[ht]
    \centering
    \includegraphics[width=1\textwidth]{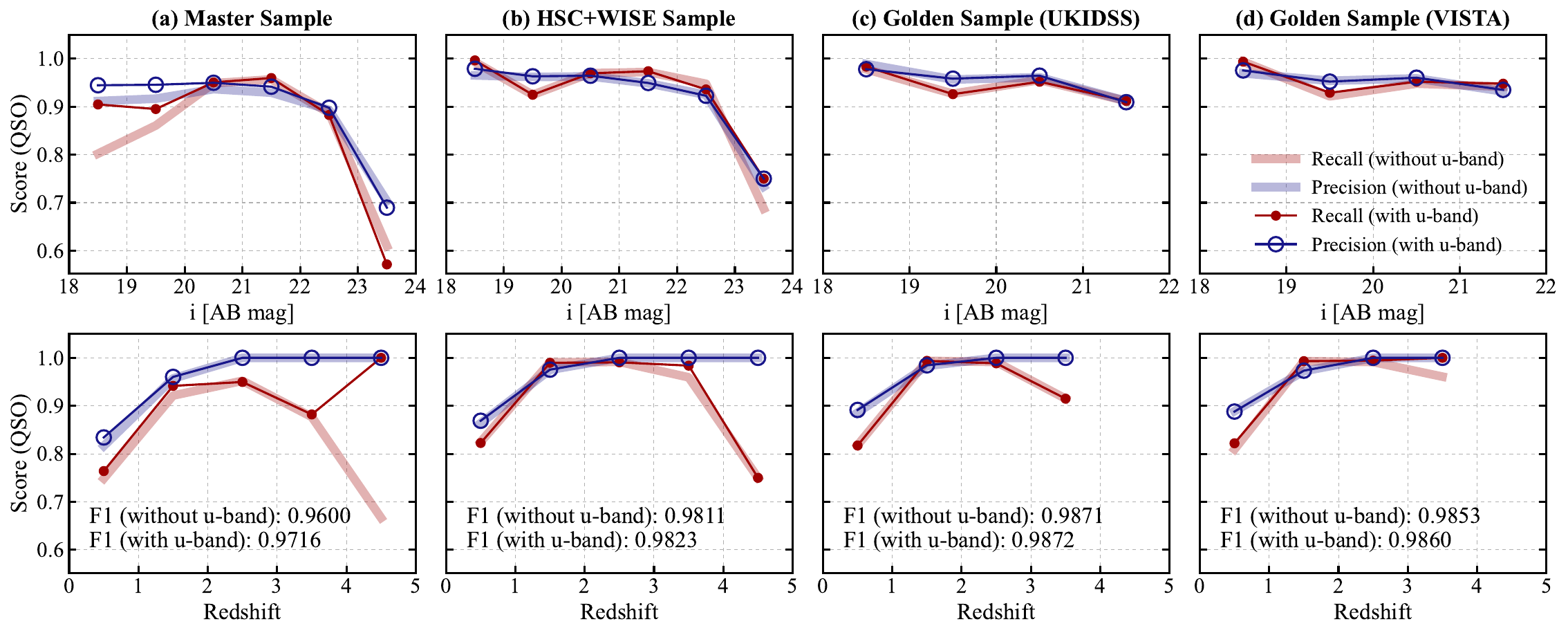}
    \caption{
    Comparison of quasar selection performance between classifiers with and without $u$-band data, evaluated across different magnitude and redshift bins for four parent samples. For each test, the training, validation, and test sets are drawn from the same random split of the corresponding parent sample. Red and blue lines denote the recall and precision scores of quasars, respectively, while thicker and thinner lines indicate classifiers without and with $u$-band data, respectively.
    }
    \label{fig:utest_bins}
\end{figure*}

\begin{figure*}[ht]
    \centering
    \includegraphics[width=1\textwidth]{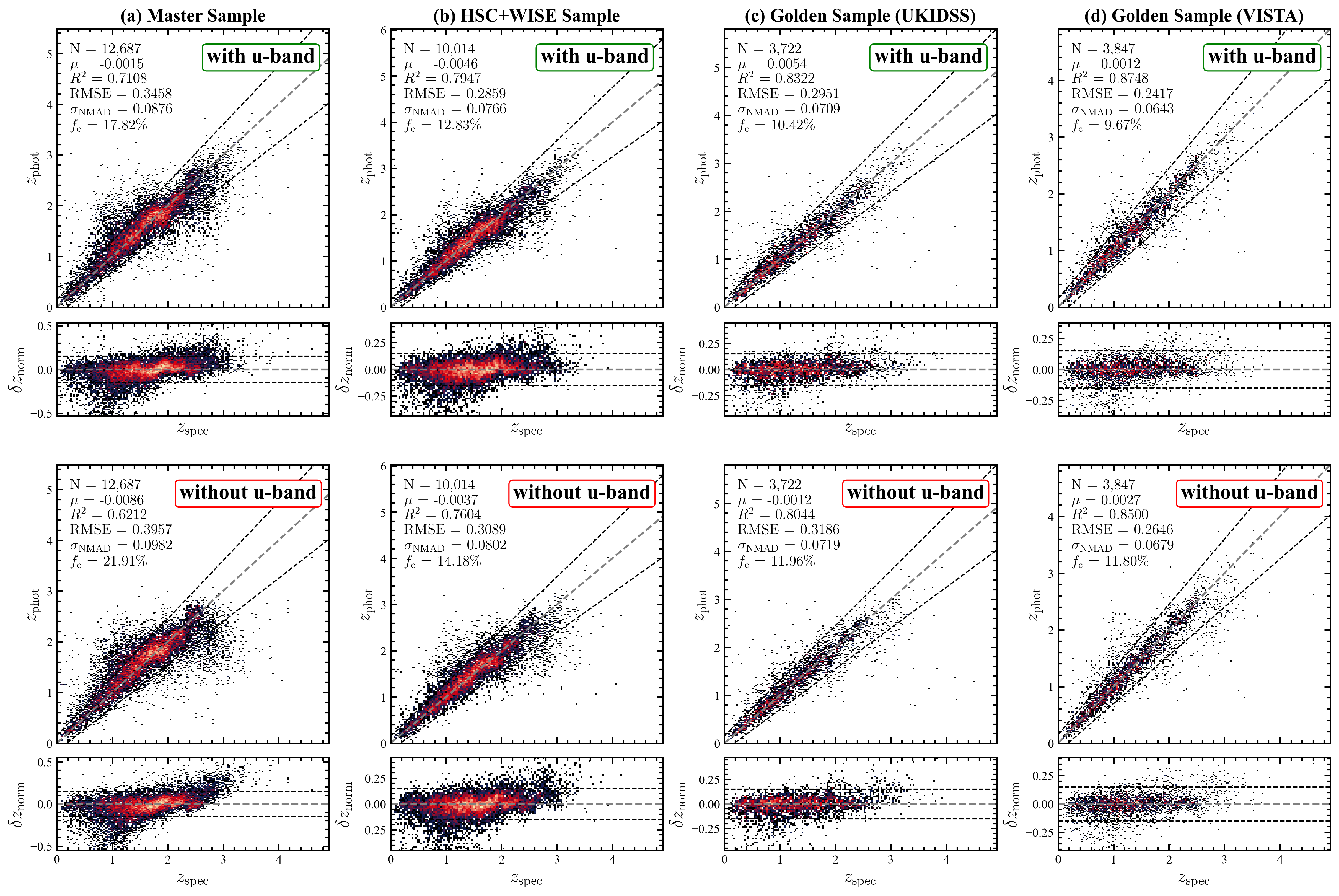}
    \caption{
    Comparison of photometric redshifts estimated by regressors with and without $u$-band data. The upper panels show results with $u$-band features, while the lower panels show results without. All symbols are as defined in Figure \ref{fig:photoz}.
    }
    \label{fig:utest_photoz}
\end{figure*}

\subsection{Evaluation of Photometric Redshift Accuracy}
Several metrics are employed to evaluate the performance of photometric redshift (photo-$z$) estimation. Assume a test set of $n$ sources, where $z_i$ denotes the spectroscopic redshift of the $i$-th source and $\hat{z}_i$ is its estimated photo-$z$ predicted by the machine learning regressor. Let $\delta z_i = z_i - \hat{z}_i$ be the residual for each source. $\bar z_i$ denotes the mean spectroscopic redshift of the test set. The performance metrics are defined as:

\begin{equation}
    \mu = \frac{1}{n} \sum_{i=1}^n (z_i - \hat{z}_i) \,,
\end{equation}

\begin{equation}
    R^2 = 1 - \frac{\sum_{i=1}^n (z_i - \hat{z}_i)^2}{\sum_{i=1}^n (z_i - \bar z_i)^2} \,,
\end{equation}

\begin{equation}
    \mathrm{RMSE} = \sqrt{\frac{1}{n} \sum_{i=1}^n (z_i - \hat{z}_i)^2} \,,
\end{equation}

\begin{equation}
    \sigma_{\mathrm{NMAD}} = 1.48 \times \mathrm{median} \left( \left| \frac{\delta z_i - \mathrm{median}(\delta z_i)}{1 + z_i} \right| \right) \,,
\end{equation}

\begin{equation}
    f_c = \frac{1}{n} \times \mathrm{count} \left( \left| \frac{\delta z_i}{1 + z_i} \right| > 0.15 \right) \,.
\end{equation}

The symbol $\mu$ represents the mean of the residuals for a given test set. The $R^2$ score measures how well the predicted redshifts match the spectroscopic redshifts by comparing the residual variance to the variance of the true values. The RMSE quantifies the overall prediction error. The $\sigma_{\mathrm{NMAD}}$ provides a robust estimate of the spread of normalized errors, less sensitive to outliers than the standard deviation. The catastrophic outlier fraction $f_c$ indicates the proportion of sources whose normalized error exceeds 0.15, representing significant prediction failures. 

Each input dataset with reliable spectroscopic redshifts is randomly divided into training and test sets with a 9:1 ratio. The final bagging XGBoost ensemble model is trained on the training set. All magnitudes used in the photometric redshift estimation have been corrected for Galactic extinction, as described in Section~\ref{subsubsec:extinction}.

Using only the five broad HSC bands, the model achieves a catastrophic outlier fraction of $f_c = 22.03\%$ and a normalized median absolute deviation of $\sigma_{\mathrm{NMAD}} \approx 0.087$. Incorporating WISE photometry significantly reduces the catastrophic failure rate to $f_c = 13.37\%$, demonstrating the importance of mid-IR bands. For the two Golden Samples, which include near-IR data and thus better sample the quasar SEDs, the performance improves further, lowering outlier rates to approximately $10\%$ and $\sigma_{\mathrm{NMAD}}$ to 0.06 (see Figure~\ref{fig:photoz}).

Table~\ref{tab:photoz} shows that the bagging XGBoost regressor consistently outperforms a single XGBoost regressor across all datasets. Furthermore, the bagging XGBoost model provides an model uncertainty estimate, \texttt{z\_ML\_std}, for each predicted photo-$z$. Applying a threshold of \texttt{z\_ML\_std} $< 0.1$ to exclude sources with high model uncertainty further enhances performance on the test set.

\section{Impact of $u$-Band Photometry on Classification Performance and Photo-$z$ Estimation} \label{sec:uband}
We investigate the impact of adding $u$-band data to the input features on classifier performance and photo-$z$ estimation. The $u$-band data used in this study are derived from the SCUSS survey, introduced in Section \ref{subsubsec:SCUSS}, selected for its depth and extensive coverage within the HSC wide fields. All sources in the HSC parent samples are cross-matched with SCUSS $u$-band photometry. These subsamples are randomly split into training-validation and test sets in a 4:1 ratio. The test sets are used exclusively to evaluate classifier performance, while the training-validation sets are further divided into separate training and validation sets using the same 4:1 ratio. The ML models, both with and without $u$-band features, utilize the same training, validation, and test sets from each parent sample.

The features used for comparison are identical to those described in Section \ref{subsec:classification} for the main selection process. The new feature sets are based on the original features, with the addition of u-band data. These additional features include u-band photometry (PSF, CModel, and $2''$ aperture photometry), the $u-g$ color, and the u-band extendedness ($u_{\mathrm{PSF}} - u_{\mathrm{CModel}}$). The XGBoost settings for u-band evaluations are slightly adjusted compared to the strategy outlined in Section \ref{subsec:classification}. Since XGBoost hyperparameter tuning is sensitive to the distribution of the training set and feature selection, we adopt the default regularization hyperparameters and apply the early stopping technique (\texttt{num\_boost\_round=1000} and \texttt{early\_stopping\_rounds=50}) to mitigate overfitting and maximize model performance. The precision and recall rates of quasars, as a function of HSC $i$ magnitudes and redshift, are shown in Figure \ref{fig:utest_bins}. Adding u-band features significantly improves performance at the bright end of the master sample but results in only limited improvements for the other three samples. Furthermore, the addition of u-band features shows minimal improvement across each redshift bin.

The impact of adding $u$-band data on photo-$z$ estimation is shown in Figure \ref{fig:utest_photoz}. For the master sample, incorporating u-band data reduces the outlier rate from $21.91\%$ to $17.82\%$, and improves $\sigma_{\mathrm{NMAD}}$ by $\sim 0.01$, as it better constrains the Lyman-$\alpha$ break at high redshift. However, the improvement is limited for the other three samples.

\section{Summary and Future Work} \label{sec:summary}

\begin{deluxetable*}{ccccc}  
\tablecaption{Format of the QHSC Quasar Candidate Catalog \label{tab:QHSC_table_format}} 
\tablewidth{0pt}  
\tablehead{
\colhead{Column} & \colhead{Name} & \colhead{Type} & \colhead{Unit} & \colhead{Description}
}
\startdata
1 & \texttt{object\_id} & int & -- & Unique HSC object ID (64-bit integer) \\
2 & \texttt{ra} & float & deg & Right Ascension (J2000) from HSC \\
3 & \texttt{dec} & float & deg & Declination (J2000) from HSC \\
4 & \texttt{pQSO} & float & -- & Probability of the object being a quasar \\
5 & \texttt{z\_ML} & float & -- & Photometric redshift estimated by the bagging XGBoost model \\
6 & \texttt{z\_ML\_std} & float & -- & Model Uncertainty of the estimated photometric redshift \\
7 & \texttt{ebv} & float & mag & Galactic extinction $E(B{-}V)$ from the SFD dust map, scaled by a factor of 0.86 \\
-- & \texttt{x\_image\_flag} & int & -- & HSC image quality flag for band $x$, where $x \in \{\mathrm{g}, \mathrm{r}, \mathrm{i}, \mathrm{z}, \mathrm{y}\}$ \\
-- & \texttt{flag\_x\_aper} & int & -- & Aperture photometry flag in HSC band $x$ \\
-- & \texttt{mag\_x\_aper} & float & mag & Aperture magnitude in HSC band $x$ \\
-- & \texttt{magerr\_x\_aper} & float & mag & Error of aperture magnitude in HSC band $x$ \\
-- & \texttt{flag\_x\_psf} & int & -- & PSF photometry flag in HSC band $x$ \\
-- & \texttt{mag\_x\_psf} & float & mag & PSF magnitude in HSC band $x$ \\
-- & \texttt{magerr\_x\_psf} & float & mag & Error of PSF magnitude in HSC band $x$ \\
-- & \texttt{flag\_x\_cmodel} & int & -- & CModel photometry flag in HSC band $x$ \\
-- & \texttt{mag\_x\_cmodel} & float & mag & CModel magnitude in HSC band $x$ \\
-- & \texttt{magerr\_x\_cmodel} & float & mag & Error of CModel magnitude in HSC band $x$ \\
\hline
\multicolumn{5}{c}{\textbf{QHSC HSC+WISE}} \\
\hline
58 & \texttt{name\_catwise} & string & -- & CatWISE2020 source name \\
59 & \texttt{mag\_W1} & float & mag & W1 magnitude from profile-fit photometry with motion correction \\
60 & \texttt{magerr\_W1} & float & mag & Error of W1 magnitude \\
61 & \texttt{mag\_W2} & float & mag & W2 magnitude from profile-fit photometry with motion correction \\
62 & \texttt{magerr\_W2} & float & mag & Error of W2 magnitude \\
\hline
\multicolumn{5}{c}{\textbf{QHSC Golden (UKIDSS)}} \\
\hline
63 & \texttt{id\_ukidss} & string & -- & Unique UKIDSS source ID \\
64 & \texttt{ukidss\_survey} & string & -- & UKIDSS sub-survey name (e.g., LAS, DXS, or UDS) \\
65 & \texttt{mag\_J} & float & mag & UKIDSS point source H aperture corrected mag ($2$ aperture diameter) \\
66 & \texttt{magerr\_J} & float & mag & Error of J-band magnitude from UKIDSS \\
67 & \texttt{mag\_H} & float & mag & UKIDSS point source H aperture corrected mag ($2$ aperture diameter) \\
68 & \texttt{magerr\_H} & float & mag & Error of H-band magnitude from UKIDSS \\
69 & \texttt{mag\_K} & float & mag & UKIDSS point source K aperture corrected mag ($2$ aperture diameter) \\
70 & \texttt{magerr\_K} & float & mag & Error of K-band magnitude from UKIDSS \\
\hline
\multicolumn{5}{c}{\textbf{QHSC Golden (VISTA)}} \\
\hline
63 & \texttt{id\_vista} & string & -- & Unique VISTA source ID \\
64 & \texttt{vista\_survey} & string & -- & VISTA survey name (e.g., VIKING, VHS, VIDEO, or UltraVISTA) \\
65 & \texttt{mag\_J} & float & mag & VISTA point source J aperture corrected mag ($2$ aperture diameter) \\
66 & \texttt{magerr\_J} & float & mag & Error of J-band magnitude from VISTA \\
67 & \texttt{mag\_H} & float & mag & VISTA point source H aperture corrected mag ($2$ aperture diameter) \\
68 & \texttt{magerr\_H} & float & mag & Error of H-band magnitude from VISTA \\
69 & \texttt{mag\_Ks} & float & mag & VISTA point source Ks aperture corrected mag ($2$ aperture diameter) \\
70 & \texttt{magerr\_Ks} & float & mag & Error of Ks-band magnitude from VISTA \\
\enddata
\tablecomments{All magnitudes are in the AB system.}
\end{deluxetable*}

The Subaru HSC-SSP Wide Survey provides deep imaging $\sim 1470~\deg^2$, reaching an $i$-band depth of $\sim 26$, enabling the construction of a large and highly complete sample of quasar candidates. By incorporating additional near-infrared and mid-infrared photometric data from the UKIDSS, VISTA and WISE surveys, contamination from stars and galaxies can be effectively reduced, although this comes at the cost of reduced sample size and shallower sample depth. Based on these photometric datasets, we perform quasar candidate classification, estimate their photo-$z$, and assess the improvements in both classification and redshift estimation with the inclusion of external $u$-band data. The main results are summarized as follows:

\begin{enumerate}
    \item We apply the XGBoost algorithm to perform quasar classification for each of the four parent samples, resulting in 1,184,574 (master), 371,777 (HSC+WISE), 87,460 (GoldenU), and 120,572 (GoldenV) quasar candidates (see Section \ref{sec:ml_selection}). Among those candidates, the pure subsamples ($\pqso > 0.95$) contain 188,430 (master), 154,307 (HSC+WISE), 55,938 (GoldenU), and 72,696 (GoldenV) sources in total. These four catalogs are stored as separate FITS extensions within the QHSC catalog file. A detailed description of the catalog format is provided in Table~\ref{tab:QHSC_table_format}. The QHSC catalog is publicly available at the Zenodo repository\footnote{\url{https://doi.org/10.5281/zenodo.17515028}}.
    
    \item We construct three external test sets from deep, magnitude-limited spectroscopic surveys (HETDEX, VVDS, and zCOSMOS-bright) to evaluate the performance of our classifiers under realistic conditions (see Sections~\ref{subsec:classifiers_evaluation} and~\ref{subsec:bin_evaluation}). The results show that the master sample achieves high completeness ($>85\%$), and that the incorporation of WISE mid-infrared data significantly improves both completeness and purity, with both metrics exceeding $90\%$.
    
    \item We compare the HSC+WISE sample with the Golden Samples and also test the effect of adding SCUSS $u$-band data. We find that both near infrared data (when mid infrared data are already included) and $u$-band data offer only limited improvements in quasar selection and photometric redshift estimation in our work.
    
    \item We employ an ensemble learning approach with XGBoost as the base learner combined with bagging (i.e. bagging XGBoost regressor), achieving better performance than a single XGBoost regressor. Model uncertainty ($\mathrm{z\_ML\_std}$) is estimated by the standard deviation across base learners. Subsamples selected with lower $\mathrm{z\_ML\_std}$ exhibit significantly higher photo-$z$ accuracy.
\end{enumerate}

To achieve maximum completeness, we retain all candidates for which the quasar class has the highest probability, i.e., those satisfying $\pqso > \mathrm{p_{GALAXY}}$ and $\pqso > \mathrm{p_{STAR}}$. By applying a threshold cut on $\pqso$, users can extract subsamples with higher purity. The HSC $i$-band magnitude and redshift distributions for the full candidate catalog, as well as for the pure subsample ($\pqso > 0.95$), are shown in Figure~\ref{fig:QHSC_mag_z_hist}. The median $i$-band magnitudes for the full candidate samples are 22.03 (master), 21.58 (HSC+WISE), 20.22 (GoldenU), and 20.79 (GoldenV); for the pure subsamples, the medians are 21.28 (master), 20.81 (HSC+WISE), 20.06 (GoldenU), and 20.53 (GoldenV).

The QHSC quasar candidate catalog is one of the deepest machine-learning selected quasar catalogs publicly available, and spanning a wide redshift range ($z_\mathrm{phot} < 6$). This catalog represents a valuable resource for further constraining the faint-end of the quasar luminosity function, as well as for identifying faint, lensed quasar systems. Furthermore, we are developing a systematic method for identifying high-redshift quasar candidates using machine learning techniques applied to the HSC survey data,  and the result will be reported in a future work.

Our results offer a reliable evaluation of classifier performance and demonstrate the efficacy of combining multi-wavelength data with advanced ML techniques for quasar selection and photometric redshift estimation, particularly in faint datasets. This work lays the foundation for quasar selection in future large-scale deep surveys, including those of LSST \citep{2019ApJ...873..111I}, Euclid \citep{2025A&A...697A...1E}, and the Chinese Space Station Telescope \citep{2011SSPMA..41.1441Z,2018cosp...42E3821Z}.

\begin{acknowledgments}
We thank the referee for constructive comments that improved the manuscript. We also thank Mara Salvato, Jing Xu, Vibhore Negi, and Dingyi Zhao for valuable discussions.
We thank the support of the National Science Foundation of China (grant No. 12133001).
This work was supported by the High-performance Computing Platform of Peking University.

The Hyper Suprime-Cam (HSC) collaboration includes the astronomical communities of Japan and Taiwan, and Princeton University. The HSC instrumentation and software were developed by the National Astronomical Observatory of Japan (NAOJ), the Kavli Institute for the Physics and Mathematics of the Universe (Kavli IPMU), the University of Tokyo, the High Energy Accelerator Research Organization (KEK), the Academia Sinica Institute for Astronomy and Astrophysics in Taiwan (ASIAA), and Princeton University. Funding was contributed by the FIRST program from the Japanese Cabinet Office, the Ministry of Education, Culture, Sports, Science and Technology (MEXT), the Japan Society for the Promotion of Science (JSPS), Japan Science and Technology Agency (JST), the Toray Science Foundation, NAOJ, Kavli IPMU, KEK, ASIAA, and Princeton University. 
This paper makes use of software developed for Vera C. Rubin Observatory. We thank the Rubin Observatory for making their code available as free software at \url{http://pipelines.lsst.io/}.
This paper is based on data collected at the Subaru Telescope and retrieved from the HSC data archive system, which is operated by the Subaru Telescope and Astronomy Data Center (ADC) at NAOJ. Data analysis was in part carried out with the cooperation of Center for Computational Astrophysics (CfCA), NAOJ. We are honored and grateful for the opportunity of observing the Universe from Maunakea, which has the cultural, historical and natural significance in Hawaii. 
This publication makes use of data products from the Wide-field Infrared Survey Explorer, which is a joint project of the University of California, Los Angeles, and the Jet Propulsion Laboratory/California Institute of Technology, funded by the National Aeronautics and Space Administration.
The UKIDSS source tables were provided by the Wide Field Astronomy Unit (WFAU) of the University of Edinburgh and originate from UKIDSS DR11 of the WFCAM science archive \citep{2008MNRAS.384..637H}. The UKIDSS project is defined in \citet{2007MNRAS.379.1599L}. UKIDSS uses the UKIRT Wide Field Camera \citep[WFCAM; ][]{2007A&A...467..777C} and a photometric system described in \citet{2006MNRAS.367..454H}.
The VISTA Hemisphere Survey data products served at Astro Data Lab are based on observations collected at the European Organisation for Astronomical Research in the Southern Hemisphere under ESO programme 179.A-2010, and/or data products created thereof.
This publication has made use of data from the VIKING survey from VISTA at the ESO Paranal Observatory, programme ID 179.A-2004. Data processing has been contributed by the VISTA Data Flow System at CASU, Cambridge and WFAU, Edinburgh.
Based on data products from observations made with ESO Telescopes at the La Silla Paranal Observatory as part of the VISTA Deep Extragalactic Observations (VIDEO) survey, under programme ID 179.A-2006 (PI: Jarvis).
Based on data products from observations made with ESO Telescopes at the La Silla Paranal Observatory under ESO programme ID 179.A-2005 and on data products produced by TERAPIX and the Cambridge Astronomy Survey Unit on behalf of the UltraVISTA consortium.
This research uses data from the VIMOS VLT Deep Survey, obtained from the VVDS database operated by Cesam, Laboratoire d'Astrophysique de Marseille, France.
Based on zCOSMOS observations carried out using the Very Large Telescope at the ESO Paranal Observatory under Programme ID: LP175.A-0839.
Funding for the Sloan Digital Sky Survey V has been provided by the Alfred P. Sloan Foundation, the Heising-Simons Foundation, the National Science Foundation, and the Participating Institutions. SDSS acknowledges support and resources from the Center for High-Performance Computing at the University of Utah. SDSS telescopes are located at Apache Point Observatory, funded by the Astrophysical Research Consortium and operated by New Mexico State University, and at Las Campanas Observatory, operated by the Carnegie Institution for Science. The SDSS web site is \url{www.sdss.org}.
SDSS is managed by the Astrophysical Research Consortium for the Participating Institutions of the SDSS Collaboration, including Caltech, The Carnegie Institution for Science, Chilean National Time Allocation Committee (CNTAC) ratified researchers, The Flatiron Institute, the Gotham Participation Group, Harvard University, Heidelberg University, The Johns Hopkins University, L'Ecole polytechnique f\'{e}d\'{e}rale de Lausanne (EPFL), Leibniz-Institut f\"{u}r Astrophysik Potsdam (AIP), Max-Planck-Institut f\"{u}r Astronomie (MPIA Heidelberg), Max-Planck-Institut f\"{u}r Extraterrestrische Physik (MPE), Nanjing University, National Astronomical Observatories of China (NAOC), New Mexico State University, The Ohio State University, Pennsylvania State University, Smithsonian Astrophysical Observatory, Space Telescope Science Institute (STScI), the Stellar Astrophysics Participation Group, Universidad Nacional Aut\'{o}noma de M\'{e}xico, University of Arizona, University of Colorado Boulder, University of Illinois at Urbana-Champaign, University of Toronto, University of Utah, University of Virginia, Yale University, and Yunnan University.
This research used data obtained with the Dark Energy Spectroscopic Instrument (DESI). DESI construction and operations is managed by the Lawrence Berkeley National Laboratory. This material is based upon work supported by the U.S. Department of Energy, Office of Science, Office of High-Energy Physics, under Contract No. DE–AC02–05CH11231, and by the National Energy Research Scientific Computing Center, a DOE Office of Science User Facility under the same contract. Additional support for DESI was provided by the U.S. National Science Foundation (NSF), Division of Astronomical Sciences under Contract No. AST-0950945 to the NSF’s National Optical-Infrared Astronomy Research Laboratory; the Science and Technology Facilities Council of the United Kingdom; the Gordon and Betty Moore Foundation; the Heising-Simons Foundation; the French Alternative Energies and Atomic Energy Commission (CEA); the National Council of Humanities, Science and Technology of Mexico (CONAHCYT); the Ministry of Science and Innovation of Spain (MICINN), and by the DESI Member Institutions: www.desi.lbl.gov/collaborating-institutions. The DESI collaboration is honored to be permitted to conduct scientific research on I’oligam Du’ag (Kitt Peak), a mountain with particular significance to the Tohono O’odham Nation. Any opinions, findings, and conclusions or recommendations expressed in this material are those of the author(s) and do not necessarily reflect the views of the U.S. National Science Foundation, the U.S. Department of Energy, or any of the listed funding agencies.
This work made use of data from the Hobby-Eberly Telescope Dark Energy Experiment (HETDEX). 
HETDEX is led by the University of Texas at Austin McDonald Observatory and Department of Astronomy with participation from the Ludwig-Maximilians-Universit\"at M\"unchen, Max-Planck-Institut f\"ur Extraterrestrische Physik (MPE), Leibniz-Institut f\"ur Astrophysik Potsdam (AIP), Texas A\&M University, Pennsylvania State University, Institut f\"ur Astrophysik G\"ottingen, The University of Oxford, Max-Planck-Institut f\"ur Astrophysik (MPA), The University of Tokyo and Missouri University of Science and Technology.
Observations for HETDEX were obtained with the Hobby-Eberly Telescope (HET), which is a joint project of the University of Texas at Austin, the Pennsylvania State University, Ludwig-Maximilians-Universit\"at M\"unchen, and Georg-August-Universit\"at G\"ottingen. The HET is named in honor of its principal benefactors, William P.~Hobby and Robert E.~Eberly. The Visible Integral-field Replicable Unit Spectrograph (VIRUS) was used for HETDEX observations. VIRUS is a joint project of the University of Texas at Austin, Leibniz-Institut f\"ur Astrophysik Potsdam (AIP), Texas A\&M University, Max-Planck-Institut f\"ur Extraterrestrische Physik (MPE), Ludwig-Maximilians-Universit\"at M\"unchen, Pennsylvania State University, Institut f\"ur Astrophysik G\"ottingen, University of Oxford, and the Max-Planck-Institut f\"ur Astrophysik (MPA).
The authors acknowledge the Texas Advanced Computing Center (TACC) at The University of Texas at Austin for providing high performance computing, visualization, and storage resources that have contributed to the research results reported within this paper. URL: \url{http://www.tacc.utexas.edu}
Funding for HETDEX has been provided by the partner institutions, the National Science Foundation, the State of Texas, the US Air Force, and by generous support from private individuals and foundations.
This research has made use of the Astrophysics Data System, funded by NASA under Cooperative Agreement 80NSSC21M00561.
This research uses services or data provided by the Astro Data Lab, which is part of the Community Science and Data Center (CSDC) Program of NSF NOIRLab. NOIRLab is operated by the Association of Universities for Research in Astronomy (AURA), Inc. under a cooperative agreement with the U.S. National Science Foundation.

\end{acknowledgments}

\facilities{Subaru, WISE, UKIRT, ESO:VISTA, Sloan, Mayall}
\software{
    astrokit\citep{2025zndo..15321579Z}, 
    astropy\citep{2013A&A...558A..33A,2018AJ....156..123A,2022ApJ...935..167A}, 
    pandas\citep{2022zndo...3509134T},
    STILTS\citep{2006ASPC..351..666T},
    TOPCAT\citep{2005ASPC..347...29T},
    scikit-learn\citep{2011JMLR...12.2825P}, 
    optuna\citep{2019arXiv190710902A}, 
    XGBoost\citep{2016arXiv160302754C}
}


\bibliography{export-bibtex}{}
\bibliographystyle{aasjournalv7}


\end{CJK*}
\end{document}